\documentclass[prb,aps,showpacs,citeautoscript, superscriptaddress,amsmath,amssymb,floatfix,,twocolumn,dvipsnames]{revtex4-1}
\usepackage{graphicx}
\usepackage{xcolor}
\usepackage{ytableau}
\usepackage[colorlinks,bookmarks=false,citecolor=blue,linkcolor=red,urlcolor=black]{hyperref}
\usepackage{amsfonts}
\usepackage{amsmath}
\usepackage{times}
\usepackage{amssymb}
\usepackage{changes}
\usepackage{blkarray}
\usepackage{relsize}

\ytableausetup{smalltableaux} 

\setcitestyle{super}

\usepackage{color}
\definecolor{lightpink}{RGB}{255, 187, 218}
\definecolor{lightblue}{RGB}{130, 215, 255}
\definecolor{lightgrey}{RGB}{220, 220, 220}
\definecolor{lightred}{RGB}{255, 155, 155}
\definecolor{lightteal}{RGB}{143, 255, 243}
\definecolor{lightindigo}{RGB}{191, 181, 255}
\definecolor{lightgreen}{RGB}{185, 255, 179}
\definecolor{lightpurple}{RGB}{237, 179, 255}

\definecolor{c1}{RGB}{144,12,63}
\definecolor{c2}{RGB}{199,0,57}
\definecolor{c3}{RGB}{255,87,51}
\definecolor{c4}{RGB}{255,195,0}
\definecolor{c5}{RGB}{218,247,166}

\begin{document}

%\title{Density matrix renormalization group simulations of $SU(N)$ Heisenberg chains using standard Young tableaux and comparison with finite size Bethe Ansatz}
\title{DMRG simulations of $SU(N)$ Heisenberg chains using standard Young tableaux: fundamental representation and comparison with finite-size Bethe ansatz}

\date{\today} 
\author{Pierre Nataf}
\affiliation{Institute of Theoretical Physics, \'Ecole Polytechnique F\'ed\'erale de Lausanne (EPFL), CH-1015 Lausanne, Switzerland}
\affiliation{Univ. Grenoble Alpes, CEA INAC-PHELIQS, F-38000, Grenoble, France}
\author{Fr\'ed\'eric Mila}
\affiliation{Institute of Theoretical Physics, \'Ecole Polytechnique F\'ed\'erale de Lausanne (EPFL), CH-1015 Lausanne, Switzerland}

\begin{abstract}
We develop an efficient method to perform density matrix renormalization group simulations of the $SU(N)$ Heisenberg chain with open boundary conditions taking full advantage of the $SU(N)$ symmetry of the problem. This method is an extension of the method previously developed for exact diagonalizations and relies on a systematic use of the basis of standard Young tableaux. Concentrating on the model with the fundamental representation at each site (i.e. one particle per site in the fermionic formulation), we have benchmarked our results for the ground state energy up to $N=8$ and up to 420 sites by comparing them with Bethe ansatz results on open chains, for which we have derived and solved the Bethe ansatz equations. The agreement for the ground state energy is excellent for SU(3) (12 digits). It decreases with $N$, but it is still satisfactory for $N=8$ (6 digits). Central charges $c$ are also extracted from the entanglement entropy using the Calabrese-Cardy formula, and agree with the theoretical values expected from the $SU(N)_1$ Wess-Zumino-Witten CFTs.
\end{abstract}

\maketitle

 \section{Introduction}
Continuous progress in ultracold atoms have allowed experimentalists to engineer more and more advanced models of strongly correlated systems. 
 In particular, degenerate gases of strontium and ytterbium loaded in optical lattices
have been used to simulate the SU(N) Fermi-Hubbard models \cite{WuPRL2003,Honerkamp2004,Cazalilla2009,gorshkov2010,takahashi2012,Pagano2014,Scazza2014,Zhang2014},
with $N$ as large as $10$. 
Very recently, the experimental realization of the $SU(N)$ Mott insulating phase was reported\cite{Hofrichter_2016}.
This class of lattice models is a unique playground for strongly correlated systems as it encompasses a wide variety of quantum ground states with different physical properties. These properties usually depend on the geometry of the lattice (one- dimensional chain, two-dimensional bipartite or frustrated lattice), the number of {\it colors} (i.e., the value of N), and the local $SU(N)$ symmetry of the wave function (the irreducible representation of $SU(N)$, or more conveniently the $SU(N)$ {\it irrep}).

From a theoretical point of view, the study of these  lattice models is very challenging. Apart from a few $SU(N)$ models which admit a Bethe ansatz solution (e.g. the $SU(N)$ Heisenberg chain with the fundamental irrep at each site), there is no reliable analytical approach to investigate these models. One then needs to use numerical methods, but they are often of limited efficiency, especially when $N$ is larger than $2$ ($N=2$ corresponds to the familiar Heisenberg spin models). 
 The main reason is the dramatic exponential increase of the dimension of the Hilbert Space, which scales as $h^{N_s}$, where $N_s$ is the number of sites, and $h$ is the dimension of the local Hilbert space. In the case of one particle per site, in the fundamental irrep, $h=N$, and it is clear that the numerical simulations will become more and more difficult as $N$ increases. 
In that respect, Quantum Monte Carlo simulations are the exception since a large local Hilbert space is not a major obstacle, but minus sign free simulations are only possible in very specifc cases\cite{frischmuth,assaad2005,bonnes,Messio2012,lang2013}.

One strategy to overcome such a difficulty, and to effectively reduce the relevant number of degrees of freedom in those models, is to {\it implement} the full $SU(N)$ symmetry in the simulations.  Let us take a practical example: in the case of the antiferromagnetic Heisenberg SU(10) model on a 20-site lattice with one particle per site, the implementation of the $SU(10)$ symmetry in an exact diagonaization (ED) algorithm allows one to work directly in the $16796-$dimensional singlet subspace to get the ground state. This dimension is several orders of magnitude smaller than the dimension of the full Hilbert space (around $10^{20}$). This is of course a huge gain since the problem has been reduced to the diagonalization of a matrix of dimension $16796 \times 16796$.
However, writing the matrix representing the Hamiltonian in the singlet subspace is in general  a difficult task since it makes use of the Clebsch-Gordan 
coefficients (CGCs), which can only be calculated with an algorithm whose complexity also increases with $N$ \cite{alex2011}.

An alternative approach has been developed for exact diagonalizations. It is based on the {\it orthogonal representation of the symmetric group} \cite{nataf2014,PRBNataf2016,Wan2017}.
A basis for such a representation is provided by the set of standard Young tableaux (SYT), and Alfred Young found out  in the 1930's \cite{Young6,rutherford} very simple rules (the {\it Young's rules}) to build the matrices representing the elements of the symmetric group in such a basis. The usefulness of the algebraic tools developed to study the representation of the {\it symmetric} group in the investigation of the Heisenberg $SU(N)$ models is a direct consequence of the {\it Schur-Weyl duality} \cite{weyl}, which establishes many relations between the representations of the $SU(N)$ groups and the representations of the symmetric groups. A familiar manifestation of the connections between the $SU(N)$ symmetry group and the permutations appears explicitly when one rewrites the  Heisenberg interaction $\vec{S}_1\cdot\vec{S}_2$ between two spins $1/2$ as a function of the permutation $P_{1,2}$ as $\vec{S}_1 \cdot \vec{S}_2=(1/2)P_{1,2}-1/4$.
Thus, it is somehow natural to use the theory of the representation of the symmetric group to investigate $SU(N)$ models.
In particular, the SYTs and the Young's rules allows one to write the matrix representing the $SU(N)$ Heisenberg models in an $SU(N)$ invariant sector (corresponding to an $SU(N)$ irrep) in a very simple manner. It is not obvious however to which extent Young's rules and SYTs can also be used to implement the $SU(N)$ symmetry in other numerical methods, such as the density matrix renormalization group (DMRG)\cite{white1993}. The purpose of this paper is to proceed with such a development.

Note that implementing non-abelian symmetries (in particular the $SU(N)$ symmetry) in a DMRG algorithm has already been achieved.
For the SU(2) case, several attempts have been successful\cite{daul,ostlund,sakamoto,mcculloch2007,mcculloch2002,Sierra_1996}. In most cases, they rely in some way on CGCs.
Interestingly, some approaches naturally lead to a formulation where the {\it colors} are factorized out: for instance, thanks to the Wigner Eckart theorem, one approach naturally leads to the use of a {\it reduced basis} \cite{mcculloch2002}, or of a {\it reduced multiplet space} \cite{weichselbaum2012}.
Through a mapping which transforms a vertex Hamiltonian into an Interaction-Round-a-Face (IRF) Hamiltonian, the DMRG approach from Nishino and Sierra \cite{Sierra_1996} employs as a relevant  {\it reduced basis} the set of {\it Bratelli diagrams}. In the case of SU(N), these latter are indeed nothing but a graphical way to represent the SYTs.
If the set of SYTs is indeed completely equivalent to the different representations of the  {\it reduced basis} used in these approaches,
the Young's rules and the orthogonal representation of the symmetric group have unfortunately not been taken full advantage of, while it dramatically reduces the complexity of the calculation of some group theory coefficients, which are otherwise based on CGCs.  In particular, it allows one to completely shortcut the difficult calculation of the $3n-j$ symbols which are useful to calculate the reduced matrix elements appearing in the Wigner-Eckart theorem, and which are necessary to calculate the matrix elements between the states of the  {\it reduced basis}.
Improving the computation of such coefficients is indeed not just an optimization problem.
As pointed out by A. Weichselbaum, whose  general framework does not apply only to $SU(N)$ invariant groups but to other non-abelian groups as well \cite{weichselbaum2012}, developing a strategy to shortcut the sum rules and contraction of complex networks of CGC space is of crucial importance to implement symmetries with rank (i.e. the value of $N$) higher than $3$.
The use of the Young's rules and SYTs is a way to implement this strategy, and it has enabled us to study the $SU(8)$ Heisenberg chain almost as easily as the $SU(3)$ Heisenberg chain.
In the following, we develop this method.

In section \ref{method}, we describe our procedure, which is based on the original formulation of DMRG by White.
Such a framework was the most natural one to use in a relatively simple manner the Young's rules.
We apply our formalism to the Heisenberg $SU(N)$ chain with one particle per site in the fundamental irrep and with open boundary conditions (OBC), but many of the concepts developed here can be generalized to other models.
We put more emphasis on the parts where Young's rules have revealed to be particularly convenient to make the group theory calculations needed for the computation.
Then, in order to have proper numerical values to benchmark our code, we explicitly write and solve the Bethe ansatz equations for the $SU(N)$ chain with OBC in section \ref{exact}.
In section \ref{results}, we compare the results from our DMRG code to the exact results for SU(3) up to SU(8), calculate the central charges for chain of several hundreds of sites, and demonstrate some good agreement with the expected $SU(N)_1$ CFT behavior.
Then, conclusions and perspectives are drawn.
Finally, some results about the orthogonal representation of the symmetric group are reported in an Appendix. Among other things, it is shown that this representation makes the determination of the IRF weights needed in the DMRG formulation of Nishino and Sierra \cite{Sierra_1996} for SU(2) essentially trivial.
 
\section{DMRG with implementation of the full $SU(N)$ symmetry}
\subsection{Structure of the DMRG code}
\label{method}
\begin{figure} 
\centerline{\includegraphics[width=1\linewidth]{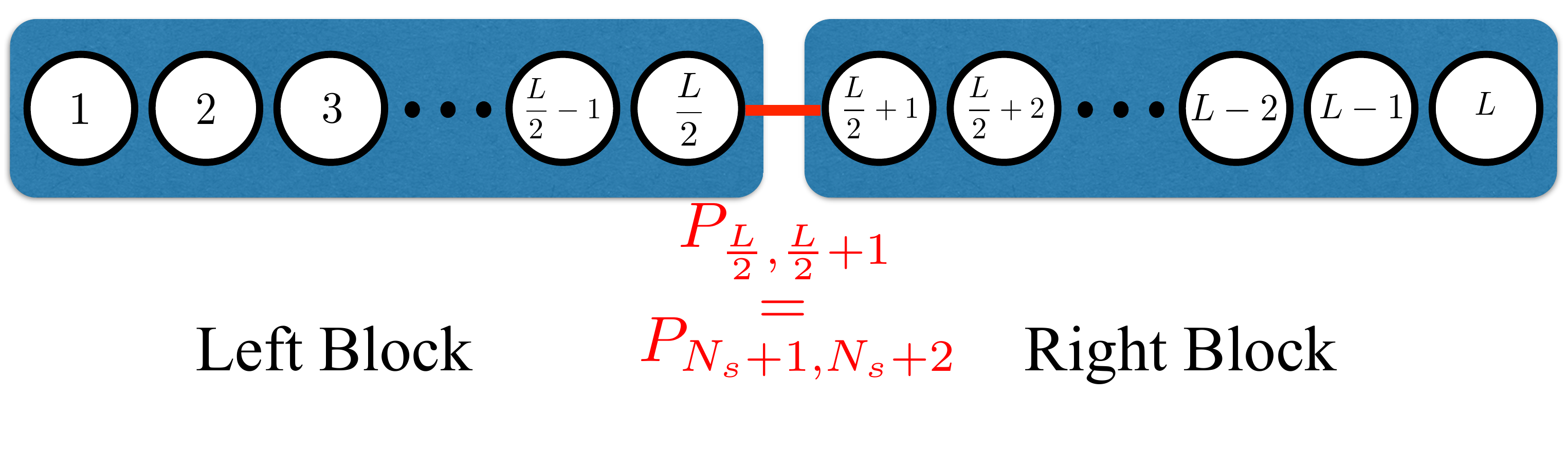}}
\caption{\label{indexing_chain} Sketch of the  indexing of the $L$-site chain ($L$ even) split into two parts: the left block and the right block.
Each block has length $L/2=N_s+1$.
The interaction between the two blocks is the permutation between the last site of the left block and the first site of the right block, i.e. $P_{\frac{L}{2},\frac{L}{2}+1}=P_{N_s+1,N_s+2}$.
}
\end{figure}

We present below a description of the DMRG algorithm with implementation of the full $SU(N)$ symmetry for the case of the antiferromagnetic Heisenberg $SU(N)$ chain
with one particle per site, i.e. with the fundamental irrep at each site. 
Most of the ideas described below can be generalized  to Hamiltonians with other irreps of SU(N), and/or other one-dimensional geometries such as ladders.
We will concentrate on the stages where the use of SYT reveals both convenient and powerful for the calculation of some group theory coefficients which are necessary for the implementation of the $SU(N)$ symmetry.  The calculation of these coefficients becomes intractable without SYTs when N increases (larger than $4$ typically).
The infinite-size DMRG algorithm we present below is based on White's original idea : the $L$-site chain is split into two half-chains of size $L/2$ (the {\it left} block and the {\it  right} block, as shown in Fig. \ref{indexing_chain}), and one lets them grow by adding one site at a time. To prevent the dimension of the Hilbert space on both blocks from becoming too large, one then applies the density matrix truncation in order to keep a number of states on both blocks not higher than $m$, which is one of the input parameters of the algorithm.
Once the desired length of the chain is reached, the infinite size part of the DMRG is finished and one starts the finite-size part which consists in performing some sweeps which allows one to gain some accuracy on the ground state energy and to determine for instance the profile of the entanglement entropy. 
We will focus below on the infinite-size part of the DMRG since most of the issues raised by the implementation of the $SU(N)$ symmetry in this part appear in a very similar way in the
 finite-size part. Before describing one step of the infinite DMRG, let us explain the truncation scheme we have chosen.  In addition to the usual density matrix truncation which involves the selection for the left (or the right) block of the $m$ states of largest eigenvalues in the diagonalization of the reduced density matrix, we perform a truncation on the $SU(N)$ irreps as well.
 
\begin{figure} 
\centerline{\includegraphics[width=1\linewidth]{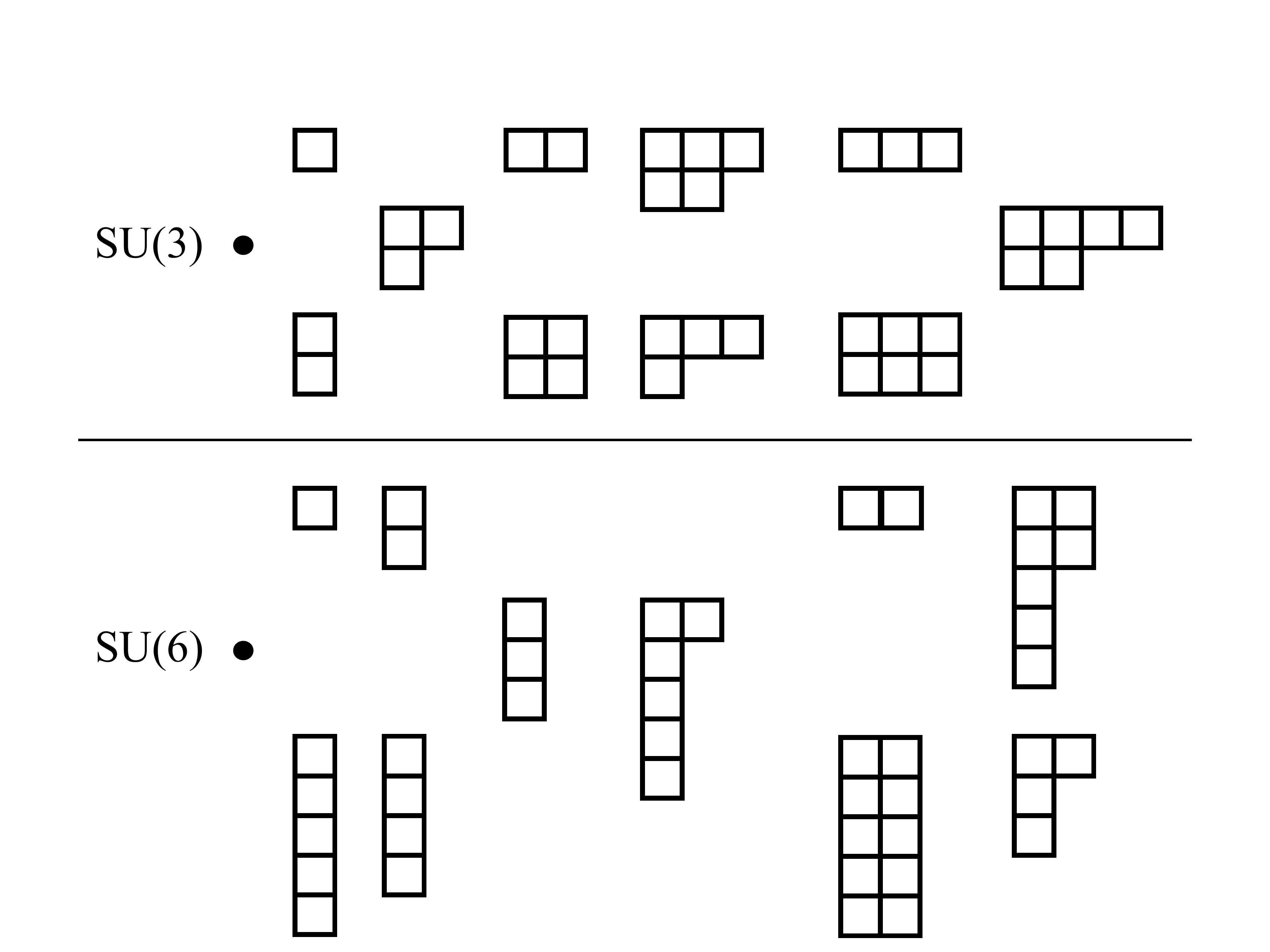}}
\caption{\label{liste_irrep}First eleven irreps of SU(3) (top) and SU(6) (bottom), ordered from left to right according to the ascending quadratic casimir. Pairs of conjugate irreps are put one above the other since they have the same quadratic casimir. The singlet irrep is drawn with a dot. The truncation scheme is such that we consider only the states belonging to the first $M$ irreps, where $M$ is an input parameter of the DMRG algorithm (see text for details).}
\end{figure}

 Only the states living in the $M$ irreps of lowest quadratic Casimir will be taken into account, where $M$ is an input parameter of the algorithm.
 This is well adapted to antiferromagnetic problems since the low energy states live in those irreps, and it allows one to calculate a {\bf finite} number of group theory coefficients (cf paragraph \ref{subduction_coeff}) needed for the computation {\it once for all} before the simulation starts. Typically, for $M$ around one or two hundreds, the accuracy is already very good for $m=1000$ as we will see in section \ref{results}. Such a list for SU(3) and for SU(6) with $M=11$ is shown in Fig. (\ref{liste_irrep}).
 The quadratic Casimir $C_2$ depends on the shape of an $SU(N)$ irrep $\alpha=[\alpha_1,\alpha_2,\cdots,\alpha_k]$, with $\alpha_i$ the length of the row $i$ of the shape and $k$ the number of rows of the shape as \cite{schellekens}:
 \begin{align}
 C_2=\frac{1}{2}\Big{\{}n(N-\frac{n}{N})+\sum_{i=1}^k\alpha_i^2-\sum_{j=1}^{j=\alpha_1}c_j^2\Big{\}},
 \end{align}
where the $c_j$ ($j=1,..,\alpha_1$) are the lengths of the columns, and $n$ is the number of boxes of the irrep.

\subsection{Description of one step of the infinite DMRG}
We explain here how to pass from the stage where both the {\it left} and the {\it right} blocks have $N_s$ sites to the stage where they both have $N_s+1$ sites.
We focus on the left block (which is perfectly mirror symmetric to the right block in the infinite size DMRG), and we assume that we have kept in memory $m_{N_s}\leq m$ states from the previous stage, each of them belonging to a given $SU(N)$ symmetry sector labelled by one Young Tableau (YT)  or {\it shape} $\alpha$ with $N_s$ boxes. Before going further, let us make a remark: there are different conventions to represent the same $SU(N)$ irrep $\alpha$: with or without the columns of $N$ boxes; for instance, for SU(3), we have 
\begin{align}
\ydiagram{3,2,1}\equiv \ydiagram{2,1}.
\end{align}
%Most of the time,  we will prefer to label the SU(N) irreps characterizing the $N_s$-particles wave-functions with Young tableaux having $N_s$ boxes. 
%\subsubsection{Legacy from the previous stage}
\subsubsection{Selection of states for the current step}
\begin{figure*} 
\centerline{\includegraphics[width=0.8\linewidth]{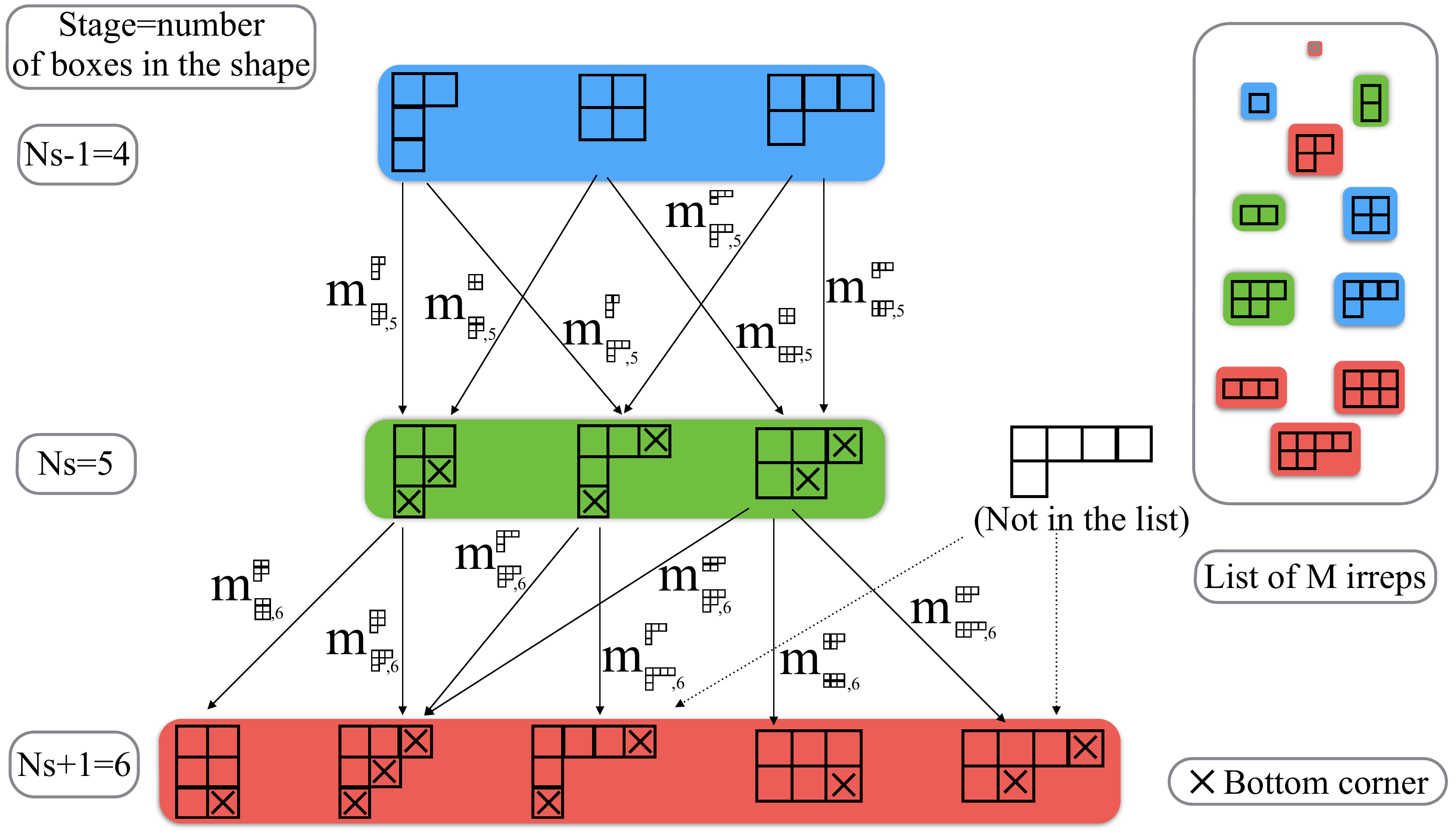}}
\caption{\label{selection} Scheme of the selection of states for $SU(3)$ and $N_s=5$. Each state belongs to a sector labelled by a Young Tableau, or shape, which is present in the list of $M$ irreps.
Such a list is shown on the right for $M=11$. Each shape $\beta$ at stage $N_s+1$ has $n_{acs}(\beta)$ ascendant shapes $\alpha \in \{ \alpha^{\beta}_1,\alpha^{\beta}_2,..,\alpha^{\beta}_{n_{asc}(\beta)} \}$ coming from the stage $N_s$, which can be obtained from $\beta$ by deleting one bottom corner (see also Fig  \ref{schema_bottom_corner} for a definition). The sector associated to $\beta$ is spanned by a set of states built from the states coming from the previous stage, and living in the sectors labelled by  $\alpha^{\beta}_k$ for $k=1...n_{acs}(\beta)$. The {\it genealogy} of each state (up to the {\it grand-parent} level), and the {\it Young rules} (cf section \ref{young_rep} for a review) are the unique ingredients needed to build the whole algorithm (see text for details). }
\end{figure*}

Each Young tableau of $N_s$ boxes to be considered should be present in the list of th $M$ YTs. In Fig. \ref{selection}, where $N_s=5$, the selected YT for stage $N_s$ are shown in green. The first two are allowed although they have only two boxes because one can add a column with three boxes to reach $N_s=5$ boxes. There are $M_{N_s}\leq M$ such shapes.
For each sector $\alpha$, we have kept $m^{\alpha}_{N_s}$ states that we can write as:
\begin{align}
\label{states_alpha}
\{\vert\zeta^{\alpha}_1\rangle,\vert\zeta^{\alpha}_2\rangle,\cdots,\vert\zeta^{\alpha}_{m^{\alpha}_{N_s}}\rangle\}.
\end{align}
The number of states $m^{\alpha}_{N_s}$ satisfy:
\begin{align}
\sum_{\alpha} m^{\alpha}_{N_s}=m_{N_s},
\end{align}
where the sum runs over the $M_{N_s}$ shapes $\alpha$ with $N_s$ boxes present in the list of $M$ irreps.

The states in Eq. (\ref{states_alpha}) are the eigenstates of the $m^{\alpha}_{N_s} \times m^{\alpha}_{N_s}$ reduced density matrix $\rho^{\alpha}$ (we will show later how to calculate the reduced density matrix for a given sector $\alpha$). The corresponding positive eigenvalues are ranked from the largest to the lowest: $\{\lambda^{\alpha}_1,\lambda^{\alpha}_2,\cdots,\lambda^{\alpha}_{m^{\alpha}_{N_s}}\}$. In addition, we also assume that from the previous stage we have kept the matrices $\mathcal{H}^{\alpha}_{N_s}$, which are the matrices of the Heisenberg Hamiltonian $H_{N_s}$ with OBC for $N_s$ sites, expressed in the basis of Eq. (\ref{states_alpha}): the $(i,j)$ coefficient of $\mathcal{H}^{\alpha}_{N_s}$ is $(\mathcal{H}^{\alpha}_{N_s})_{i,j}=\langle \zeta^{\alpha}_i\vert H_{N_s}\vert\zeta^{\alpha}_j \rangle$.

%\subsubsection{Selection of states for the current step}
To add one site to the left block, we first need to identify the $M_{N_s+1}\leq M$ shapes  $\beta$ containing  $N_s+1$ boxes and belonging to the list of $M$ irreps
(They are shown in red in Fig. \ref{selection} where $N_s=5$).

\begin{figure} 
\centerline{\includegraphics[width=0.7\linewidth]{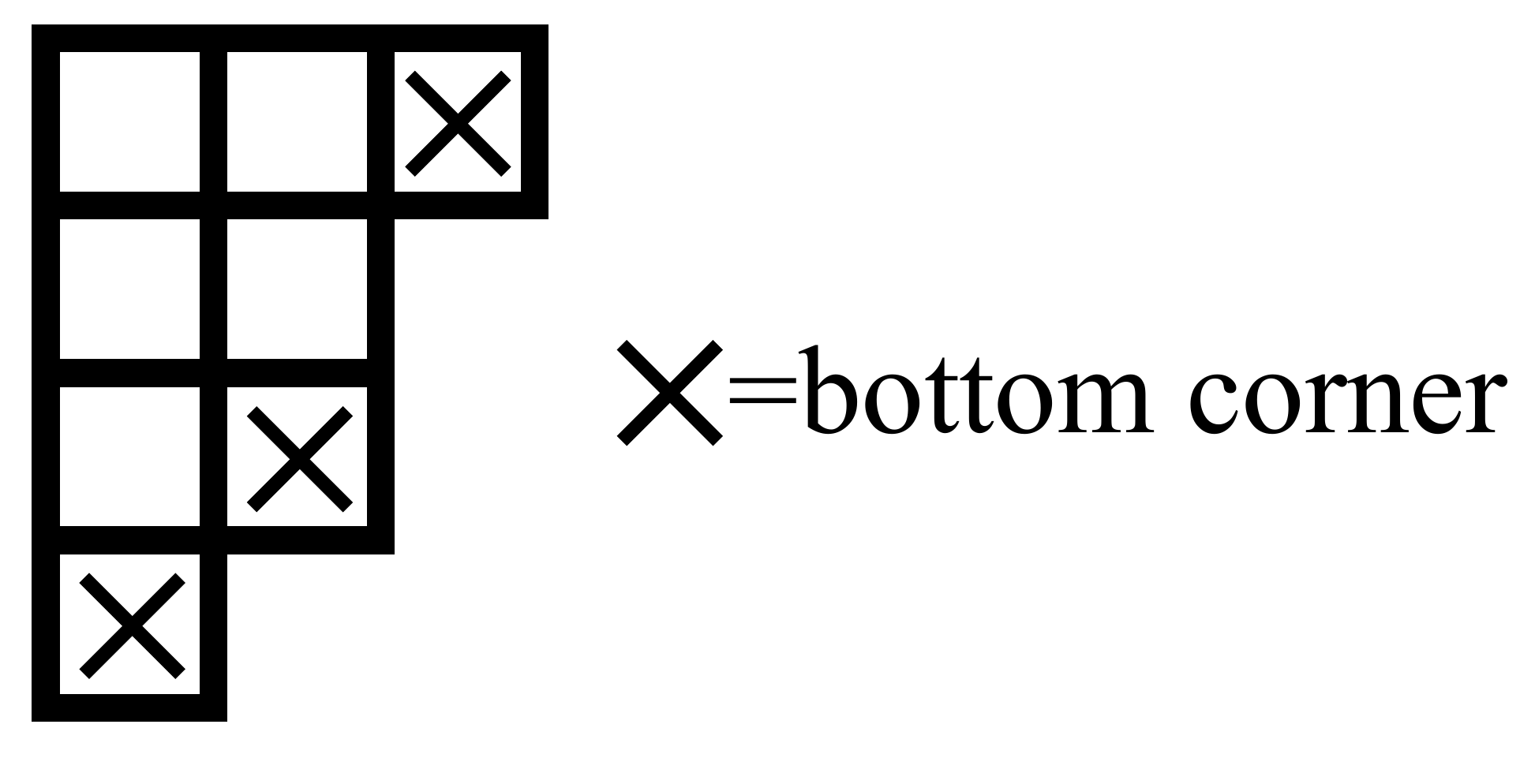}}
\caption{\label{schema_bottom_corner} For a given {\it Young tableau} $\beta=[\beta_1,\beta_2,\cdots,\beta_k]$, with $\beta_i$ the length of the row $i$ of the shape and $k$ the number of rows of the shape, (here $\beta=[3 2 2 1]$ and $k=4$), the bottom corners are the boxes where we could put the last number when we fill up the shape in the {\it standard way} (here, the last number, which is also the total number of boxes, is equal to $8$).
If we conventionally set $\beta_{k+1}=0$, those bottom corner correspond to all the rows $j$ such that $\beta_j>\beta_{j+1}$.}
\end{figure}

Then we need to select $m_{N_s+1}$ states which will live in the ensemble of $M_{N_s+1}$ sectors $\beta$. 
To perform such a selection of states (which is indeed nothing but the density matrix truncation), we proceed as follows:
for each shape $\beta$, we consider all the possible {\it ascendant} shapes $\alpha$ made of $N_s$ boxes: they are such that $\beta$ belongs to the tensor product $\alpha \otimes \ydiagram{1}$ , where $\ydiagram{1}$ is the fundamental irrep. We denote the number of ascendant shapes by $n_{asc}(\beta)$. Each ascendant of $\beta$, that one can write $\alpha_k^{\beta}$ (for $k = 1,\cdots,n_{asc}(\beta)$) is obtained by deleting one {\it bottom corner} from $\beta$ (see  Fig. \ref{schema_bottom_corner} for a definition and some examples of bottom corners).

Then, for each $\beta$, we create the list $L_{\beta}$ of eigenvalues of $\rho^{\alpha}$ for $\alpha$ an ascendant shape, i.e. $\alpha \in \{ \alpha^{\beta}_1,\alpha^{\beta}_2,..,\alpha^{\beta}_{n_{asc}(\beta)} \}$:
\begin{align}
L_{\beta}=\{\lambda^{\alpha}_q\}_{q=1,\cdots,m^{\alpha}_{N_s}}^{\alpha \in \{ \alpha^{\beta}_1,\alpha^{\beta}_2,..,\alpha^{\beta}_{n_{asc}(\beta)} \}}.
\end{align}
One then creates $L_{N_s+1}$, the union of these lists over the $M_{N_s+1}$ shapes $\beta$: $L_{N_s+1}=\underset{\beta}{\cup}L_{\beta}$.
Note that if a $N_s-$box shape $\alpha$ gives rise to several different {\it descendant} shapes when adding the fundamental irrep, its associated values $\lambda^{\alpha}_q$ (for $q=1,\cdots,m^{\alpha}_{N_s}$) will appear several times in $L_{N_s+1}$.
Then we choose the $m_{Ns+1}$  largest values in $L_{N_s+1}$, where $m_{Ns+1}=\text{Min}(m,\text{cardinal}(L_{N_s+1}))$.
The labels attached to each of the chosen $\lambda^{ \alpha^{\beta}_k}_q$ allows one to select for each sector $\beta$,  $m^{\beta}_{N_s+1}$ ascendant states.
We call $\sigma_{\beta}$ the sum of the corresponding eigenvalues.
 Each selected state comes from one of the $n_{asc}(\beta)$ ascendant shape $\alpha^{\beta}_k$. We finally denote by $m^{\alpha}_{\beta, N_s+1}$ the number of selected states belonging to the sector $\beta$ (shapes with $N_s+1$ boxes) coming from the ascendant sector $\alpha$ (shapes with $N_s$ boxes), so that $m^{\beta}_{N_s+1}=\sum_{\alpha}m^{\alpha}_{\beta, N_s+1}$.
 In Fig \ref{selection}, we give a practical example to illustrate the selection we have described in this section.
Let us note that other selection schemes exist. For instance, the set of numbers $\{m^{\beta}_{N_s+1}\}_{\beta}$ can be imposed as an input parameter (like in \cite{tatsuaki}), but for a given number of states  $\sum_{\beta}m^{\beta}_{N_s+1}$, this scheme is less efficient than the protocol shown here, which automatically selects the best set $\{m^{\beta}_{N_s+1}\}_{\beta}$.

Finally, the {\it weight} which is discarded as a consequence of  the truncation is given by:
\begin{align}
\label{error}
\mathcal{W}_d^{L}=1-\sum_{\beta} g_{\beta} \sigma_{\beta},
\end{align}
In this expression, the factor $g_{\beta}=\text{dim}(\beta)/h$, where $\text{dim}(\beta)$  is the dimension of the $SU(N)$ irrep of shape $\beta$ (see section \ref{Appendix_irrep} for a way to calculate such a dimension directly from the shape $\beta$),
and where $h$ is the dimension of the local Hilbert space. In the present case, $h=\text{dim}(\Box)=N$.
The factor $g_{\beta}$ comes from the normalization condition of the reduced density matrices (see also part \ref{next_stage}) and from the enhancement of the dimension of the full Hilbert space when one site is added to the left block.

\subsubsection{Construction of the new matrices for the left block}
\label{construction_left}
In this section, we explain how to build $\mathcal{H}^{\beta}_{N_s+1}$, the matrix representing the antiferromagnetic Heisenberg Hamiltonian $H_{N_s+1}$ with OBC for $N_s+1$ sites in each sector $\beta$  (shapes with $N_s+1$ boxes).
First of all, one can write $H_{N_s+1}$ as:
\begin{align}
H_{N_s+1}=H_{N_s}+P_{N_s,N_s+1},
\end{align}
where $H_{N_s}$ is the Heisenberg Hamiltonian for $N_s$ sites with OBC, and where $P_{N_s,N_s+1}$ is the interaction term between site $N_s$ and site $N_s+1$.
The $m^{\beta}_{N_s+1} \times m^{\beta}_{N_s+1}$ matrix  $\mathcal{H}^{\beta}_{N_s+1}$ will be written as the sum of the matrix representing respectively $H_{N_s}$ and $P_{N_s,N_s+1}$ in the sector $\beta$.
We know from the previous part that the sector $\beta$ can be split into $n_{asc}(\beta)$ subsectors, each of them corresponding to an ascendant shape $\alpha^{\beta}_k$ ($k=1,\cdots,n_{asc}(\beta)$), of dimension $m^{\alpha^{\beta}_k}_{\beta, N_s+1}$.

\begin{figure} 
\centerline{\includegraphics[width=1.25\linewidth]{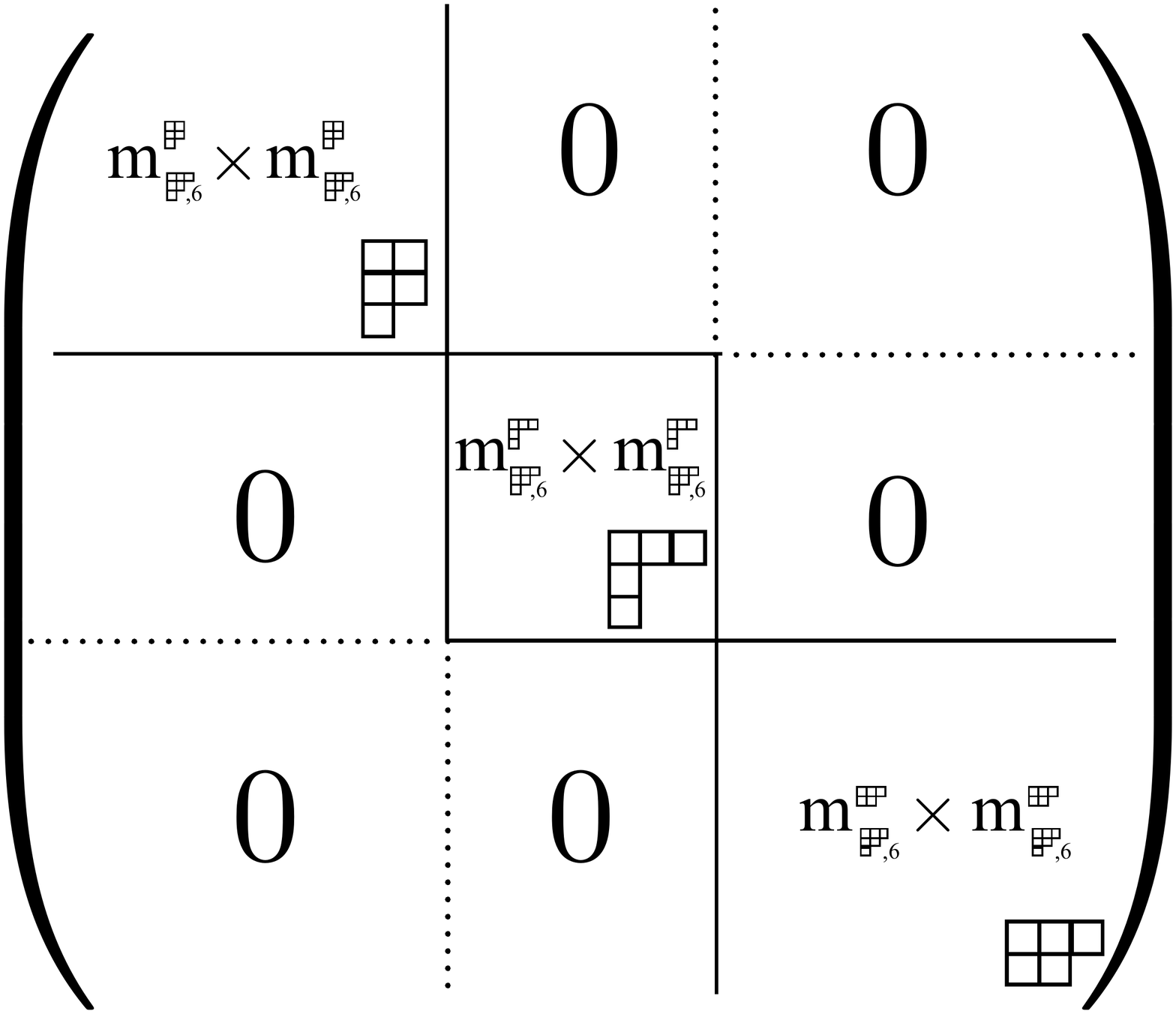}}
\caption{\label{matrix_H_Ns} Matrix $\mathcal{H}^{\beta}_{N_s}$, which represents $H_{N_s}$ (here $N_s=5$), in the sector $\beta=\ydiagram{3,2,1}$.
It is block diagonal; each block is associated to a shape $\alpha$ which is an ascendant of $\beta$ (and which is called $\alpha^{\beta}_k$ 
for $k=1,\cdots,n_{asc}(\beta)$ in the text) and has dimension $m^{\alpha}_{\beta, N_s+1} \times m^{\alpha}_{\beta, N_s+1}$. The matrix representing $H_{N_s+1}$ in the sector $\beta$
will be the sum of $\mathcal{H}^{\beta}_{N_s}$ and of another matrix $\mathcal{P}_{N_s,N_s+1}^{\beta}$ (see text for details). }
\end{figure}

Firstly, since we have kept in memory from the previous stages the matrices $\mathcal{H}^{\alpha}_{N_s}$, the construction of the matrix representing $H_{N_s}$ in the sector $\beta$ just requires the concatenation of the submatrices $(\mathcal{H}^{\alpha}_{N_s})_{i,j}$ where $1\leq i \leq m^{\alpha}_{\beta, N_s+1} ,1\leq j \leq m^{\alpha}_{\beta, N_s+1}$, for $\alpha=\alpha^{\beta}_k$ 
($k=1,\cdots,n_{asc}(\beta)$).
The matrix $\mathcal{H}^{\beta}_{N_s}$ , which represents $H_{N_s}$ in the sector $\beta$ will then be block-diagonal, each block corresponding to a subsector $\alpha^{\beta}_k$ 
($k=1,\cdots,n_{asc}(\beta)$), as illustrated in Fig \ref{matrix_H_Ns}.

Now, the construction of the matrix $\mathcal{P}_{N_s,N_s+1}^{\beta}$, which represents $P_{N_s,N_s+1}$ in the sector $\beta$, requires two kinds of ingredients:

- The set of numbers $\{m^{\alpha}_{\beta, N_s+1}\}_{\alpha \in \{ \alpha^{\beta}_1,\alpha^{\beta}_2,..,\alpha^{\beta}_{n_{asc}(\beta)} \}}$ 
and $\{m^{\chi}_{\alpha, N_s}\}_{\chi \in \{ \chi^{\alpha}_1,\chi^{\alpha}_2,..,\chi^{\alpha}_{n_{asc}(\alpha)} \}}$ for  $\alpha=\alpha^{\beta}_k$ 
($k=1,\cdots,n_{asc}(\beta)$). The shapes $\chi \in \{ \chi^{\alpha}_1,\chi^{\alpha}_2,..,\chi^{\alpha}_{n_{asc}(\alpha)} \}$ are the shapes with $N_s-1$ boxes which are ascendant of a given shape $\alpha$. 

- The set of wave-functions $\{\vert\zeta^{\alpha}_1\rangle,\vert\zeta^{\alpha}_2\rangle,\cdots,\vert\zeta^{\alpha}_{m^{\alpha}_{N_s}}\rangle\}$, $\forall \alpha=\alpha^{\beta}_k$ 
($k=1,\cdots,n_{asc}(\beta)$).

\begin{figure} 
\centerline{\includegraphics[width=1\linewidth]{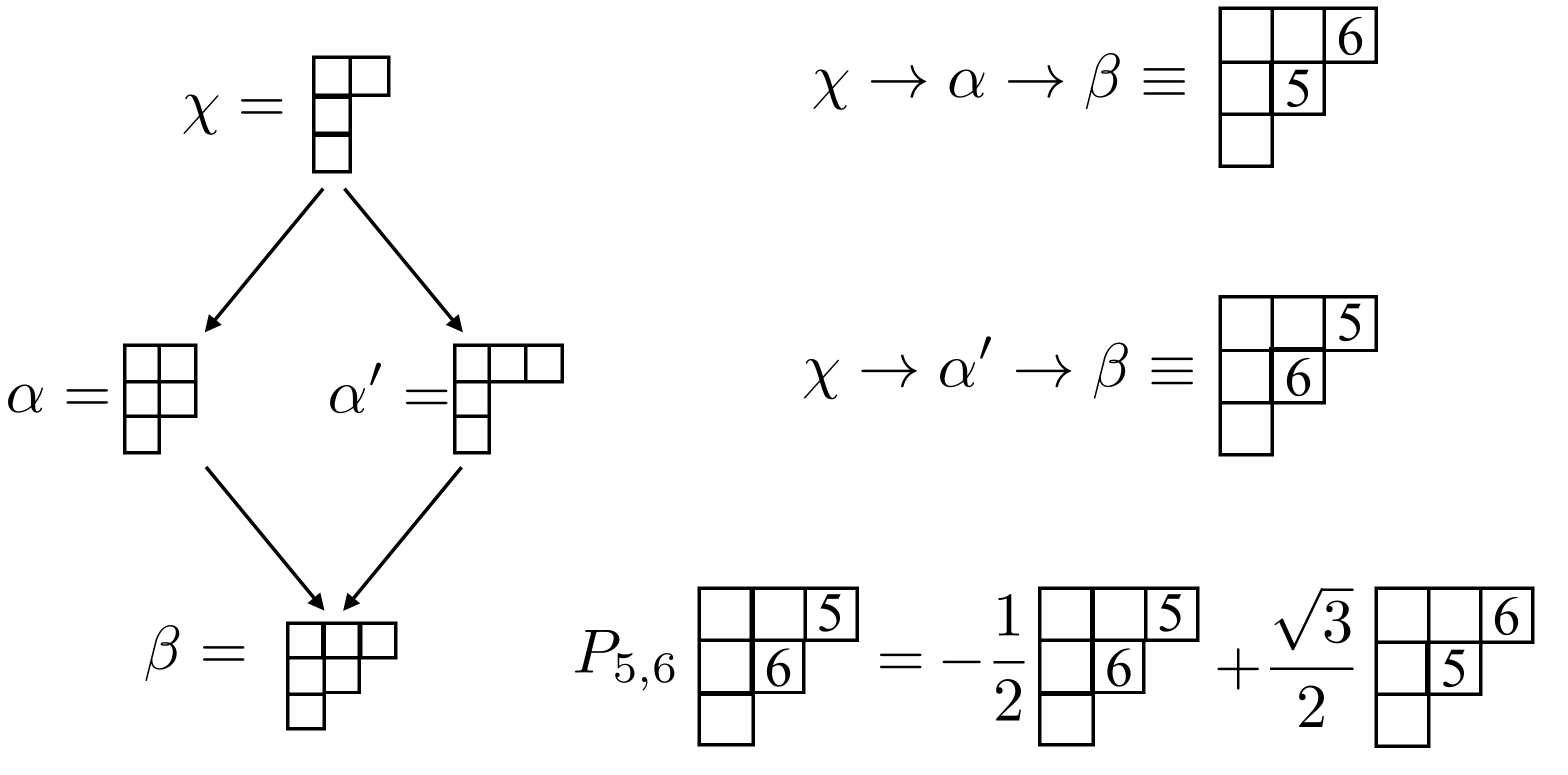}}
\caption{\label{chain} To create the matrix $\mathcal{P}_{N_s,N_s+1}^{\beta}$, which represents $P_{N_s,N_s+1}$ in the sector $\beta$ (here $\beta=\ydiagram{3,2,1}$ and $N_s=5$),
one just needs to keep track of the {\it chains} $\chi \rightarrow \alpha \rightarrow \beta$ and $\chi \rightarrow \alpha' \rightarrow \beta$, where $\alpha$ and $\alpha'$ are two ascendant shapes of $\beta$ and $\chi$ is a common ascendant shape of both $\alpha$ and $\alpha'$. Then one just uses the rules of the orthogonal representation of the symmetric group  (Young's rules) reviewed in the appendix (section \ref{young_rep}) to know the effect of $P_{N_s,N_s+1}$ on the states living in the sector $\beta$ and subsectors $\alpha$ and $\alpha'$.}
\end{figure}

The sectors $\alpha$ being split into different subsectors $\chi$, the coefficients of the corresponding wavefunctions $\vert\zeta^{\alpha}_q\rangle$ (for $q\leq m^{\alpha}_{\beta, N_s+1}$) are decomposed accordingly.
 Then, for the coefficient $\langle \zeta^{\alpha}_i\vert P_{N_s,N_s+1} \vert\zeta^{\alpha'}_j \rangle$  (with $\alpha,\alpha' \in \{ \alpha^{\beta}_1,\alpha^{\beta}_2,..,\alpha^{\beta}_{n_{asc}(\beta)} \}$ and $1\leq i \leq  m^{\alpha}_{\beta, N_s+1}$ and $1\leq j \leq  m^{\alpha'}_{\beta, N_s+1}$) to be non-zero, $\alpha$ and $\alpha'$ must have one common ascendant shape $\chi$. 
 From the shapes $\chi$, $\beta$, $\alpha$ and  $\alpha'$, the calculation of  $\langle \zeta^{\alpha}_i\vert P_{N_s,N_s+1} \vert\zeta^{\alpha'}_j \rangle$  becomes trivial thanks to the rules defining the orthogonal representation of the symmetric group (the Young's rules) that are reviewed in the appendix (in section \ref{young_rep}). This set of rules describes the action of the permutation Hamiltonian term over the basis of SYTs.
In particular, the action of $P_{N_s,N_s+1}$ just depends on the location of the bottom corner from stage $N_s-1$ to stage $N_s$ and on the location of the bottom corner from stage $N_s$ to $N_s+1$. Those informations are contained in the chains $\chi \rightarrow \alpha \rightarrow \beta$ (and $\chi \rightarrow \alpha' \rightarrow \beta$), as illustrated in Fig \ref{chain}. 

Interestingly,  in the IRF-DMRG approach that can be used for the SU(2) Heisenberg chain, the matrix elements $\langle \zeta^{\alpha}_i\vert P_{N_s,N_s+1} \vert\zeta^{\alpha'}_j \rangle$ are expressed in terms of the "Boltzmann (or IRF) weights". The calculation of these coefficients becomes trivial using the Young's rules, as we show in the appendix, in subsection \ref{irf}.

\subsubsection{Building the matrix for the superblock}
\begin{figure} 
\centerline{\includegraphics[width=1\linewidth]{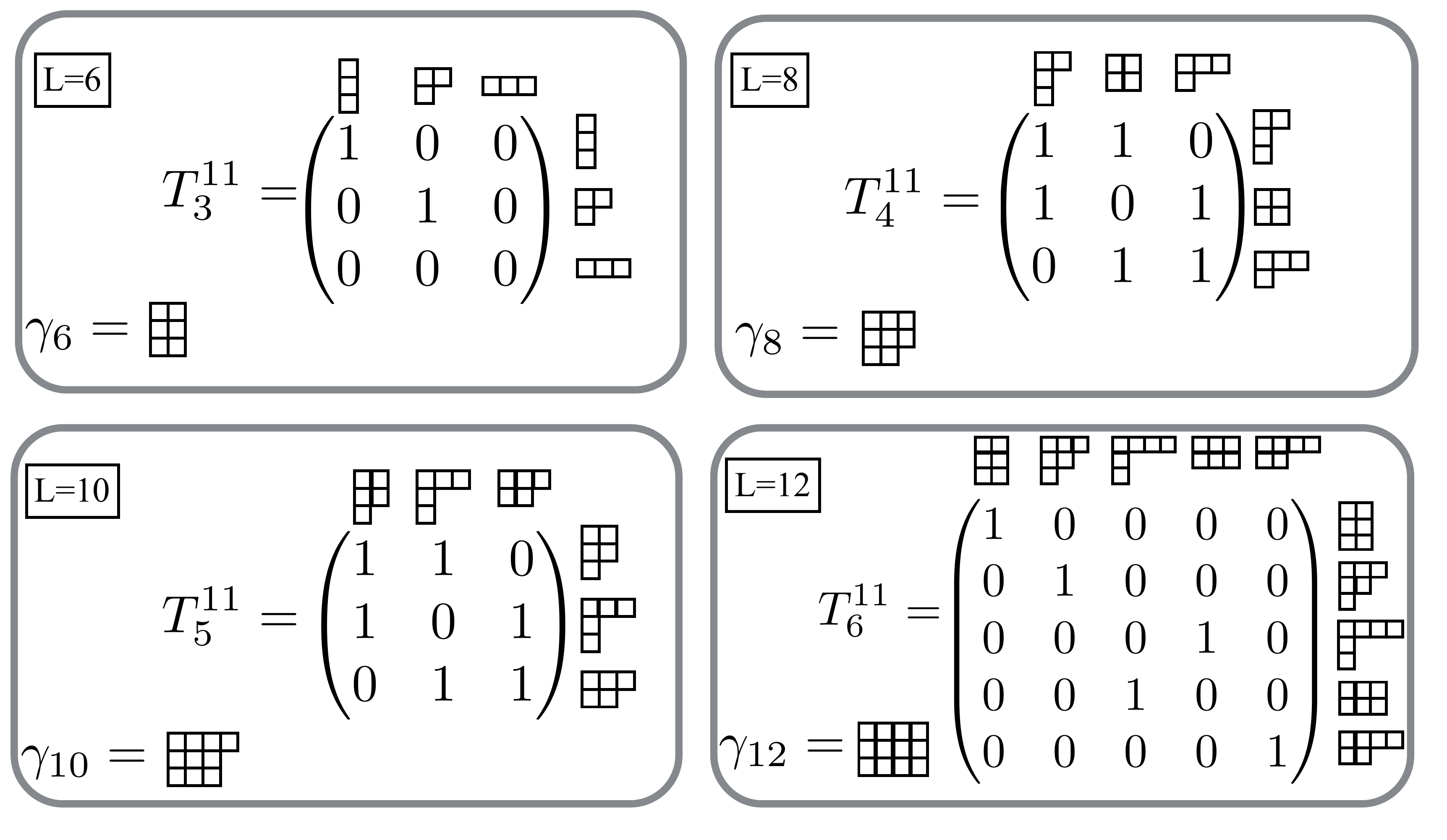}}
\caption{\label{matrix_bool} Examples of shapes $\gamma_L$ and matrices $T^{M=11}_{N_s+1}$ for different sizes: from $L=6$ (top left) to $L=12$ (bottom right), for SU(3).
The shape $\gamma_L$ is the SU(3) irrep containing the ground state for the antiferromagnetic Heisenberg SU(3) spin chain with $L$ sites with OBC and the fundamental irrep at each site.
The shapes $\beta_q$ ($q=1,\cdots,M_{N_S+1}$) which label the columns (resp. lines) of the matrices  $T^{M=11}_{N_s+1}$ represent the $SU(N)$ irreps for the left (resp. right) block made of $N_s+1=L/2$ sites.
The entries $T^{11}_{N_s+1}(q,q')=1$ if $\gamma_L \in \beta_q \otimes \beta_{q'}$ and $0$ otherwise
(see text for details). For instance,  $T^{11}_{5}(2,1)=1$ because $\ydiagram{3,1,1}\otimes \ydiagram{2,2,1}=\ydiagram{5,3,2}\oplus \ydiagram{4,3,3}$ contains $\gamma_{10}.$}
\end{figure}

So far, we have been able to create the matrices representing the antiferromagnetic Heisenberg Hamiltonian $H_{N_s+1}$ with OBC for $N_s+1$ sites in every sector $\beta$  (shapes with $N_s+1$ boxes), where $\beta$ is one of the $M_{N_s+1}$ shapes present in the list of the $M$ $SU(N)$ irreps. From now on, we index those shapes by $\beta \equiv \beta_q$ ($q=1,\cdots,M_{N_S+1}$), which are ordered according to their quadratic Casimir.
 We are going to use the matrices $\mathcal{H}^{\beta_q}_{N_s+1}$  ($q=1,\cdots,M_{N_S+1}$) for our next goal: finding the ground state $\vert GS \rangle_L$ of the antiferromagnetic Heisenberg Hamiltonian $H_{L}$ with OBC for $L=2N_s+2$ sites, which corresponds to the {\it union} of the left and the right block, as shown in Fig \ref{indexing_chain}.  To achieve this purpose, one can first build the matrix representing  $H_{L}$ on the $SU(N)$ sector where $\vert GS \rangle_{L}$ lives. Such a sector is labelled by a shape made of $L$ boxes, and is clearly model-dependent. In our case, (antiferromagnet with the fundamental irrep at each site), such a shape is equivalent to the column with $rem(L,N)$ boxes, where $rem(L,N)$ is the rest in the euclidean division of L by N. For instance, if $L=12$ and $N=3$, $\vert GS \rangle_L$ will be an $SU(3)$ singlet.
 Once such a sector $\gamma_L$ is identified, we need to create the Hilbert space of the associated {\it superblock} and to write the matrix representing $H_{L}$ in it.
 To reach this goal, one first calculates the $M_{N_s+1}\times M_{N_s+1}$  boolean (and symmetric) matrix $T^{M}_{N_s+1}$ defined for $1\leq  q,q'\leq   M_{N_s+1}$ as:
  \begin{align}
 T^{M}_{N_s+1}(q,q')&=1 \Leftrightarrow  \gamma_L \in \beta_q \otimes \beta_{q'}\nonumber \\
T^{M}_{N_s+1}(q,q')&=0 \,\,\,\text{otherwise},
 \end{align}
where $\gamma_L \in \beta_q \otimes \beta_{q'}$ means that $\gamma_L$ appears in the tensor product of the $SU(N)$ irreps $ \beta_q$ and $ \beta_{q'}$.
To perform such a tensor product, one can for instance use  the Itzykson-Nauenberg  rules \cite{itzykson}. %(or the Littelwood-Richardson rules).
In Fig. \ref{matrix_bool}, we give the matrices $T^{M}_{N_s+1}$ for SU(3), $M=11$ (cf list shown in Fig. \ref{liste_irrep}), for $N_s+1=3$ up to $N_s+1=6$.\\

We then list the $M^{GS}_{N_S+1} \leq M_{N_S+1}$ {\it relevant} shapes $\beta_k$ (for $1\leq k \leq  M_{N_S+1}$), which are such that the column $T^{M}_{N_s+1}(:,k)$ have at least one non zero entry. We write them $\tilde{\beta}_k$ (for $k=1,\cdots,M^{GS}_{N_S+1}$). For instance, for SU(3), $L=6$ and $M\geq9$, there are $M_3=3$ shapes $\beta_q$:
\begin{align}
\beta_1&=\ydiagram{1,1,1} \nonumber \\
\beta_2&=\ydiagram{2,1} \nonumber \\
\beta_3&=\ydiagram{3} \nonumber \\
\end{align}  
\\
But as shown in Fig. \ref{matrix_bool}, since the third column of $T^{11}_3$ has only vanishing entries contrary to the first two columns, $M^{GS}_{3}=2$, and the two relevant shapes
are $\tilde{\beta}_1=\ydiagram{1,1,1}$ and $\tilde{\beta}_2=\ydiagram{2,1}$. The shape $\ydiagram{3}$ does not participate into the creation of the ground state of the superblock.

The Hilbert space of the superblock on the sector $\gamma_L$ is then the tensor product of the direct sum of the sectors corresponding to the shapes $\tilde{\beta}_k$ (for $k=1,\cdots,M^{GS}_{N_S+1}$) for the left block with the same direct sum for the right block.
Consequently, its dimension is $\tilde{m}_{N_s+1}^2$, where $\tilde{m}_{N_s+1}=\sum_{k=1}^{M^{GS}_{N_S+1}} m^{\tilde{\beta}_k}_{N_s+1}$.
In practice, after a few starting steps, we observe that  $\tilde{m}_{N_s+1}=m_{N_s+1}=m$.
The ground state wave-function will thus be written as:
\begin{align}
\vert GS \rangle_L=\sum_{i,i',k,k'} \Psi^{\vert GS\rangle _L}_{i,i',k,k'} \vert \eta^{\tilde{\beta}_k}_i\rangle \otimes \vert \eta^{\tilde{\beta}_{k'}}_{i'}\rangle,
\end{align}
\\
where the sum runs over the indices $i,i',k,k'$ which are such that $1\leq k,k'\leq M^{GS}_{N_S+1}$ , $1\leq i\leq m^{\tilde{\beta}_k}_{N_s+1}$, and $1\leq i'\leq m^{\tilde{\beta}_{k'}}_{N_s+1}$,
and where the states $\vert \eta^{\tilde{\beta}_k}_i\rangle$ are the states of the sector $\tilde{\beta}_k$ (see previous section for the way we built the corresponding sector).
Now that we know what the Hilbert space for the superblock is, let us write $\mathcal{H}^{\gamma_L}_{L}$, the matrix representing  $H_{L}$ in this space.
First, we can decompose $H_{L}$ as:
\begin{align}
H_{L}=H_{N_s+1}^{\text{Left}}+H_{N_s+1}^{\text{Right}}+P_{N_s+1,N_s+2},
\end{align}
where $H_{N_s+1}^{\text{Left}}$ is the Heisenberg Hamiltonian with OBC for the left block (for site $1$ up to site $N_s+1$),
 $H_{N_s+1}^{\text{Right}}$ is the Heisenberg Hamiltonian with OBC for the right block (for site $N_s+2$ up to site $2N_s+2$),
 and $P_{N_s+1,N_s+2}$ is the interaction term between the last site of the left block (indexed by $N_s+1$), and the first site of the right block (indexed by $N_s+2$).
 We illustrate this indexing in Fig \ref{indexing_chain}.
 The calculation of the matrix $\mathcal{P}_{N_s+1,N_s+2}^{\gamma_L}$, which represents $P_{N_s+1,N_s+2}$ (on the sector $\gamma_L$), is made easy thanks to the SYTs, through the use of the {\it subduction} coefficients.
 We will focus on this task in a devoted paragraph (cf section \ref{subduction_coeff}).\\
The matrices $\mathcal{H}_{N_s+1}^{\text{Left}}$ and $\mathcal{H}_{N_s+1}^{\text{Right}}$ representing respectively $H_{N_s+1}^{\text{Left}}$ and $H_{N_s+1}^{\text{Right}}$ on the sector $\gamma_L$ of the superblock, are simply:
\begin{align}
\mathcal{H}_{N_s+1}^{\text{Left}}&=\underset{T^{M}_{N_s+1}(q,q')=1 }{\bigoplus} \mathcal{H}^{\beta_q}_{N_s+1} \otimes \mathcal{I}_{N_s+1}^{\beta_{q'}} \nonumber \\
\mathcal{H}_{N_s+1}^{\text{Right}}&=\underset{T^{M}_{N_s+1}(q',q)=1 }{\bigoplus}  \mathcal{I}_{N_s+1}^{\beta_{q'}}  \otimes \mathcal{H}^{\beta_q}_{N_s+1},
\end{align}
where $\mathcal{I}_{N_s+1}^{\beta}$ is the $m^{\beta}_{N_s+1}  \times m^{\beta}_{N_s+1} $ identity matrix on the sector labelled by the shape $\beta$ (for $\beta \in\{\tilde{\beta}_1,\tilde{\beta}_2,\cdots,\tilde{\beta}_{M^{GS}_{N_S+1}}\})$.

\subsubsection{Preparing the next stage}
\label{next_stage}
To finish this stage of the algorithm, one first calculates the ground state $\vert GS \rangle_{L}$, which is the eigenvector of minimal energy (that we will call $\mathcal{E}_L$) of $\mathcal{H}^{\gamma_L}_{L}$, using for instance the Lanczos algorithm. 
%One can calculate also excited energies and states living on $\gamma_L$, even if there are not directly needed for the algorithm, they can provide useful physical information (typically the {\it gap} in the sector $\gamma_L$).
Then, we need to calculate the reduced density matrices $\rho^{\tilde{\beta}_k}$ (for $k=1,\cdots,M^{GS}$), whose coefficients are defined (for $1\leq i,j \leq m^{\tilde{\beta}_k}_{N_s+1}$) as:
\begin{align}
\rho^{\tilde{\beta}_k}(i,j)=\frac{1}{\text{dim}(\tilde{\beta}_k)}\sum_{i',k'} (\Psi^{\vert GS\rangle _L}_{i,i',k,k'})^* \Psi^{\vert GS\rangle _L}_{j,i',k,k'},
\end{align}
where $\text{dim}(\tilde{\beta}_k)$ is the dimension of the $SU(N)$ irrep of shape $\tilde{\beta}_k$.
The factor $1/\text{dim}(\tilde{\beta}_k)$ guarantees the correct normalization of the reduced density matrices. This is completely analogous to Eqs. (68)-(69) in Ref. [\onlinecite{Sierra_1996}], which deals  with the SU(2) case.
This factor, together with the factor $g_{\beta}$ defined below Eq. (\ref{error}) of the present paper, are also consistent with the factor defined in Eq. (18) of Ref. [\onlinecite{weichselbaum2012}].

We then diagonalize the reduced density matrices to obtain the set of eigenvalues ranked from the largest to the lowest one: $\{\lambda^{\tilde{\beta}_k}_1,\lambda^{\tilde{\beta}_k}_2,\cdots,\lambda^{\tilde{\beta}_k}_{m^{{\tilde{\beta}_k}}_{N_s+1}}\}$, as well as the corresponding eigenvectors :
\begin{align}
\{\vert\zeta^{\tilde{\beta}_k}_1\rangle,\vert\zeta^{\tilde{\beta}_k}_2\rangle,\cdots,\vert\zeta^{\tilde{\beta}_k}_{m^{\tilde{\beta}_k}_{N_s+1}}\rangle\}, \label{new_basis}
\end{align}
and we perform a rotation to reexpress $\mathcal{H}^{\tilde{\beta}_k}_{N_s+1}$ in the basis shown in Eq. (\ref{new_basis}), for $k=1,\cdots,M^{GS}$. We keep in memory those quantities for the next stage.
At this step, one can also calculate the entanglement entropy $S(L)$ as:
\begin{align}
\label{entanglement_rho}
S(L)=-\sum_{k=1}^{M^{GS}_{N_S+1}} \text{dim}(\tilde{\beta}_k) \rho^{\tilde{\beta}_k} \log (\rho^{\tilde{\beta}_k}).
\end{align}
Note that the matrices representing the Heisenberg Hamiltonian in the irrelevant sectors, i.e.: 
\begin{align}
\mathcal{H}^{\beta}_{N_s+1}\,\,\text{for}\,\, \beta \not \in \{\tilde{\beta}_1, \tilde{\beta}_2,\cdots,\tilde{\beta}_{M^{GS}_{N_S+1}}\} \nonumber
\end{align}
 do not undergo any transformation at this step; furthermore,  the values $\lambda^{\beta}$ for such sectors are null. 

\subsection{Calculation of the matrix representing the interaction between the left and the right block using the permutational subduction coefficients}
 \label{subduction_coeff}

 In this section, we show how to calculate the matrix $\mathcal{P}^{\gamma_L}_{N_s+1,N_s+2}$, which represents the interblock interaction term $P_{N_s+1,N_s+2}$ on the sector $\gamma_L$.
 To simplify the notation, all along this part we will drop the indices in $\mathcal{P}^{\gamma_L}_{N_s+1,N_s+2}$ and simply call it $\mathcal{P}_{\gamma_L}$.\\
 First of all, for $\beta_{q_1},\beta_{q_2},\beta_{q_3},\beta_{q_4} \in\{\tilde{\beta}_1,\tilde{\beta}_2,\cdots,\tilde{\beta}_{M^{GS}_{N_S+1}}\}$ and for  $i_j$ such that $1 \leq i_j \leq  m^{\beta_{q_j}}_{N_s+1}$ (for $j=1,..,4$), the coefficients $\langle \eta^{\beta_{q_3}}_{i_3}  \vert \otimes\langle \eta^{\beta_{q_4}}_{i_4} \vert  \mathcal{P}_{\gamma_L} \vert \eta^{\beta_{q_1}}_{i_1}\rangle \otimes \vert \eta^{\beta_{q_2}}_{i_2}\rangle$
will be zero unless that:
\begin{align} 
T^{M}_{N_s+1}(q_1,q_2)=T^{M}_{N_s+1}(q_3,q_4)=1\,\,\,\,\,\,\,\,\,\,\,\,\text{Condition (1)}
\end{align}
 which means that both $\gamma_L\in \beta_{q_1} \otimes \beta_{q_2}$ and $\gamma_L\in \beta_{q_3} \otimes \beta_{q_4}$.
 
 Another condition is that the subsectors where the states $ \vert \eta^{\beta_{q_1}}_{i_1}\rangle$ and  $\vert \eta^{\beta_{q_3}}_{i_3}\rangle$ live (cf Fig \ref{matrix_H_Ns} for a way the sectors $\beta$ are split into different subsectors) should correspond to the same ascendant shape:
  \begin{align}
  \alpha&=\alpha^{\beta_{q_1}}_{k_1}=\alpha^{\beta_{q_3}}_{k_3}      \,\,\,\,\,\,\,\,\,\,\,\,\,\,\,\,\,\,\,\,\,\,\,\,\,\,\,\,\,\,\,\,\,\,\,\,\,\,\,\,\,\,\,\,\,\,\,\,\,\,\,\,\,\text{Condition (2)}
 % \alpha'&=\alpha^{\beta_{q_2}}_{k_2}=\alpha^{\beta_{q_4}}_{k_4}, 
  \end{align} 
   for some indices $1\leq k_j \leq n_{asc}(\beta_{q_j})$ for $j=1,3$.
   And the same condition applies for the other pair of states $ \vert\eta^{\beta_{q_2}}_{i_2}\rangle$ and $\vert\eta^{\beta_{q_4}}_{i_4}\rangle$:
   \begin{align}
  \alpha'&=\alpha^{\beta_{q_2}}_{k_2}=\alpha^{\beta_{q_4}}_{k_4}     \,\,\,\,\,\,\,\,\,\,\,\,\,\, \,\,\,\,\,\,\,\,\,\,\,\,\,\,\,\,\,\,\,\,\,\,\,\,\,\,\,\,\,\,\,\,\,\,\,\,\,\,\,\text{Condition (3)}
  \end{align} 
   for some indices $1\leq k_j \leq n_{asc}(\beta_{q_j})$ for $j=2,4$.

  The coefficients $\langle \eta^{\beta_{q_3}}_{i_3}  \vert \otimes\langle \eta^{\beta_{q_4}}_{i_4} \vert  \mathcal{P}_{\gamma_L} \vert \eta^{\beta_{q_1}}_{i_1}\rangle \otimes \vert \eta^{\beta_{q_2}}_{i_2}\rangle$ then will only depend on the four chains which determine the subsectors included in the sectors $\beta_{q_j}$ where the states $\vert \eta^{\beta_{q_j}}_{i_j}\rangle$ live
 (for $j=1,2,3,4$):
 \begin{align}
 \alpha &\rightarrow \beta_{q_1} \nonumber \\
 \alpha &\rightarrow \beta_{q_3} \nonumber \\
 \alpha' &\rightarrow \beta_{q_2} \nonumber \\
 \alpha' &\rightarrow \beta_{q_4}. \label{chain_coeff} 
 \end{align}
We can keep track of this information using the shapes $\beta_{q_j}$  and the bottom corners
which reveal which ascendant shape each state $ \vert \eta^{\beta_{q_j}}_{i_j}\rangle$ (for $j=1,2,3,4$) comes from.
For instance, for $N_s+1=5$ and $SU(3)$, as will be shown in the next subsection,
\begin{equation}
\text{\raisebox{-2.4ex}{$\mathlarger{\mathlarger{\mathlarger{\mathlarger{\mathlarger{\mathlarger{\langle}}}}}}$}}  \ytableaushort{\,\,\times,&}\otimes   \ytableaushort{\,\,\times,&}\text{\raisebox{-2.4ex}{$\mathlarger{\mathlarger{\mathlarger{\mathlarger{\mathlarger{\mathlarger{\vert}}}}}}$}}\mathcal{P}_{\text{\small $\ytableaushort{\,\,\,\,,\,\,\,,\,\,\,}$}} \text{\raisebox{-2.4ex}{$\mathlarger{\mathlarger{\mathlarger{\mathlarger{\mathlarger{\mathlarger{\vert}}}}}}$}}  \ytableaushort{\,\,,\,\,,\times}\otimes   \ytableaushort{\,\,,\,\,,\times}\text{\raisebox{-2.4ex}{$\mathlarger{\mathlarger{\mathlarger{\mathlarger{\mathlarger{\mathlarger{\rangle}}}}}}$}}\text{\raisebox{-1.4ex}{$=\sqrt{\frac{15}{16}}$}}. \label{coeff_ex}
\end{equation}
Let us as well illustrate the implications of conditions (1), (2) and (3) using this notation.
Condition (1) implies that:
\begin{equation}
\text{\raisebox{-2.4ex}{$\mathlarger{\mathlarger{\mathlarger{\mathlarger{\mathlarger{\mathlarger{\langle}}}}}}$}}  \ytableaushort{\,\,\times,&}\otimes  \ytableaushort{&\times,\,,\,} \text{\raisebox{-2.4ex}{$\mathlarger{\mathlarger{\mathlarger{\mathlarger{\mathlarger{\mathlarger{\vert}}}}}}$}}\mathcal{P}_{\text{\small $\ytableaushort{\,\,\,\,,\,\,\,,\,\,\,}$}} \text{\raisebox{-2.4ex}{$\mathlarger{\mathlarger{\mathlarger{\mathlarger{\mathlarger{\mathlarger{\vert}}}}}}$}}  \ytableaushort{\,\,\times,&} \otimes  \ytableaushort{\,\,,\,\times,\,} \text{\raisebox{-2.4ex}{$\mathlarger{\mathlarger{\mathlarger{\mathlarger{\mathlarger{\mathlarger{\rangle}}}}}}$}}\text{\raisebox{-1.4ex}{$=0$}}
\end{equation}
since $T^{11}_5(3,1)=0$ as shown in Fig. \ref{matrix_bool}.
Condition (2) leads for instance to:
\begin{equation}
\text{\raisebox{-2.4ex}{$\mathlarger{\mathlarger{\mathlarger{\mathlarger{\mathlarger{\mathlarger{\langle}}}}}}$}}  \ytableaushort{\,\,\times,&}\otimes  \ytableaushort{&\times,\,,\,} \text{\raisebox{-2.4ex}{$\mathlarger{\mathlarger{\mathlarger{\mathlarger{\mathlarger{\mathlarger{\vert}}}}}}$}}\mathcal{P}_{\text{\small $\ytableaushort{\,\,\,\,,\,\,\,,\,\,\,}$}} \text{\raisebox{-2.4ex}{$\mathlarger{\mathlarger{\mathlarger{\mathlarger{\mathlarger{\mathlarger{\vert}}}}}}$}}  \ytableaushort{\,\,,\,\times,\,}\otimes  \ytableaushort{\,\,,\,\times,\,} \text{\raisebox{-2.4ex}{$\mathlarger{\mathlarger{\mathlarger{\mathlarger{\mathlarger{\mathlarger{\rangle}}}}}}$}}\text{\raisebox{-1.4ex}{$=0,$}}
\end{equation}
while
\begin{equation}
 \text{\raisebox{-2.4ex}{$\mathlarger{\mathlarger{\mathlarger{\mathlarger{\mathlarger{\mathlarger{\langle}}}}}}$}}  \ytableaushort{\,\,\times,&}\otimes  \ytableaushort{&\times,\,,\,} \text{\raisebox{-2.4ex}{$\mathlarger{\mathlarger{\mathlarger{\mathlarger{\mathlarger{\mathlarger{\vert}}}}}}$}}\mathcal{P}_{\text{\small $\ytableaushort{\,\,\,\,,\,\,\,,\,\,\,}$}} \text{\raisebox{-2.4ex}{$\mathlarger{\mathlarger{\mathlarger{\mathlarger{\mathlarger{\mathlarger{\vert}}}}}}$}}  \ytableaushort{\,\,,\,\,,\times}\otimes  \ytableaushort{\,\,,\,\times,\,} \text{\raisebox{-2.4ex}{$\mathlarger{\mathlarger{\mathlarger{\mathlarger{\mathlarger{\mathlarger{\rangle}}}}}}$}}\text{\raisebox{-1.4ex}{$\neq 0.$}}
\end{equation}
As for condition (3), it implies for instance that:
\begin{equation}
\text{\raisebox{-2.4ex}{$\mathlarger{\mathlarger{\mathlarger{\mathlarger{\mathlarger{\mathlarger{\langle}}}}}}$}}  \ytableaushort{\,\,\times,&}\otimes  \ytableaushort{&\times,\,,\,} \text{\raisebox{-2.4ex}{$\mathlarger{\mathlarger{\mathlarger{\mathlarger{\mathlarger{\mathlarger{\vert}}}}}}$}}\mathcal{P}_{\text{\small $\ytableaushort{\,\,\,\,,\,\,\,,\,\,\,}$}} \text{\raisebox{-2.4ex}{$\mathlarger{\mathlarger{\mathlarger{\mathlarger{\mathlarger{\mathlarger{\vert}}}}}}$}}  \ytableaushort{\,\,,\,\,,\times}\otimes  \ytableaushort{\,\,,\,\,,\times} \text{\raisebox{-2.4ex}{$\mathlarger{\mathlarger{\mathlarger{\mathlarger{\mathlarger{\mathlarger{\rangle}}}}}}$}}\text{\raisebox{-1.4ex}{$=0.$}}
\end{equation}
Finally, we should keep in mind that we are dealing with $SU(N)$ irreps so the coefficients for two ensembles of shapes which are equivalent: $\{\beta_{q_1},\beta_{q_2},\beta_{q_3},\beta_{q_4}\}\equiv \{\beta'_{q_1},\beta'_{q_2},\beta'_{q_3},\beta'_{q_4}\} $ should be the same. In particular, the coefficients appearing in $\mathcal{P}_{\gamma_L}$ at stage $N_s+pN$ ($p$ integer) should be the same as the ones appearing at stage $N_s$. For instance,
\begin{equation}
\text{\raisebox{-2.4ex}{$\mathlarger{\mathlarger{\mathlarger{\mathlarger{\mathlarger{\mathlarger{\langle}}}}}}$}}  \ytableaushort{\,\,\,\times,\,&,\,}\otimes  \ytableaushort{\,&\times,\,\,,\,\,} \text{\raisebox{-2.4ex}{$\mathlarger{\mathlarger{\mathlarger{\mathlarger{\mathlarger{\mathlarger{\vert}}}}}}$}}\mathcal{P}_{\text{\small $\ytableaushort{\,\,\,\,\,\,,\,\,\,\,\,,\,\,\,\,\,}$}} \text{\raisebox{-2.4ex}{$\mathlarger{\mathlarger{\mathlarger{\mathlarger{\mathlarger{\mathlarger{\vert}}}}}}$}}  \ytableaushort{\,\,\,,\,\,\,,\,\times}\otimes  \ytableaushort{\,\,\,,\,\,\times,\,\,} \text{\raisebox{-2.4ex}{$\mathlarger{\mathlarger{\mathlarger{\mathlarger{\mathlarger{\mathlarger{\rangle}}}}}}$}} \nonumber 
\end{equation}
\begin{equation}
\text{\raisebox{-1.4ex}{$=$}}\text{\raisebox{-2.4ex}{$\mathlarger{\mathlarger{\mathlarger{\mathlarger{\mathlarger{\mathlarger{\langle}}}}}}$}}  \ytableaushort{\,\,\times,&}\otimes  \ytableaushort{&\times,\,,\,} \text{\raisebox{-2.4ex}{$\mathlarger{\mathlarger{\mathlarger{\mathlarger{\mathlarger{\mathlarger{\vert}}}}}}$}}\mathcal{P}_{\text{\small $\ytableaushort{\,\,\,\,,\,\,\,,\,\,\,}$}} \text{\raisebox{-2.4ex}{$\mathlarger{\mathlarger{\mathlarger{\mathlarger{\mathlarger{\mathlarger{\vert}}}}}}$}}  \ytableaushort{\,\,,\,\,,\times}\otimes  \ytableaushort{\,\,,\,\times,\,} \text{\raisebox{-2.4ex}{$\mathlarger{\mathlarger{\mathlarger{\mathlarger{\mathlarger{\mathlarger{\rangle}}}}}}$}}
\end{equation}

It also means that the number of such coefficients needed to implement our algorithm is finite: we calculate and store them before we actually start the simulation.
But the number of such coefficients scales like $\sim M N^5$. First, there are $M$ shapes in the list (input of the algorithm), and for each of them
there are at most $N$ shapes which allow to create the sector $\gamma_L$. For instance, for $L=8$, there are $2$ non vanishing entries per column in $T^{11}_4$. More generally this number scales as $N$. Then, for each of the four shapes appearing in the symbol, there are at most $N$ bottom corners.
For $N=8$ and $M=200$, $M N^5$ is already several millions, so we need also an efficient algorithm to calculate those coefficients.
We describe one of them  below.

 \subsubsection{Application of Chen's method to calculate the subduction coefficients}

In this section, we will show how to calculate the  coefficients $\langle \beta_{q_3},l_3  \vert \otimes\langle \beta_{q_4},l_4 \vert  \mathcal{P}_{\gamma_L} \vert \beta_{q_1},l_1\rangle \otimes \vert \beta_{q_2},l_2\rangle$, where  $\beta_{q_j}$ is a shape (i.e. $SU(N)$ irrep or Young diagram), and $l_j$ is the row of the corresponding bottom corner (which satisfies $1\leq l_j \leq N$ and 
which is enough to determine the exact location of the bottom corner inside $\beta_{q_j}$), for $j=1,2,3,4$. 
We will assume that the conditions 1, 2 and 3 stated above are satisfied.
We will describe in detail the different steps to calculate this kind of coefficients, and we will apply them to the SU(3) example introduced above (cf Eq. (\ref{coeff_ex}))  in which $\beta_{q_1}=\beta_{q_2}=[2,2,1]$, $\beta_{q_3}=\beta_{q_4}=[3,2]$, $l_1=l_2=3$, and $l_3=l_4=1$, in order to prove Eq. (\ref{coeff_ex}).
The idea is to use the SYTs and the orthogonal representation of the symmetric group to make this  kind of calculation. 

\underline{Step 1}

- We first replace the couple $(\beta_{q_1},l_1)$ by a SYT $S_1$ of the same shape, having its last number located in the bottom corner $l_1$: $(\beta_{q_1},l_1)\rightarrow S_1$. For our example, one can take for instance:
\begin{align}
\ytableaushort{\,\,,\,\,,\times} \rightarrow \ytableaushort{13,24,5}=S_1.
\end{align}  

- For the second couple $(\beta_{q_2},l_2)$, we make a slighly different replacement: we replace the couple $(\beta_2,l_2)$ by the SYT $S_2^t$ which has the shape $\beta_2$ and which is the last one in the last letter order (cf appendix, section \ref{young_rep} for a definition), and we keep in memory the bottom corner $l_2$ for the next step. Then we reindex the numbers located in $S^t_2$ to agree with the indexing in Fig \ref{indexing_chain}: $(1,2,\cdots)\rightarrow (L,L-1,\cdots)$ in order to obtain $S_2$. Thus, we have the transformation $(\beta_{q_2},l_2)\rightarrow S^t_2 \rightarrow S_2$. For our example, one has:
\begin{align}
\ytableaushort{\,\,,\,\,,\times}  \rightarrow \ytableaushort{\,\,,\,\,,\,}  \rightarrow \ytableaushort{14,25,3} \rightarrow  \begin{ytableau}
 10 & 7\\9 &6\\8 \end{ytableau}=S_2.
\end{align}  

\underline{Step 2}

  We then expand the "product" $S_1\otimes S_2$ on the shape $\gamma_L$ employing Chen's method which makes use of the 
{\it permutational subduction coefficients} (cf section 4.18.3 in \cite{chen}).
One must find a linear superposition of SYTs of shape $\gamma_L$ having the same properties of "internal symmetries" between particles as the 
product $S_1\otimes S_2$. For instance, the product:
\begin{align}
\ytableaushort{13,24,5}\otimes \begin{ytableau}
 10 & 7\\9 &6\\8 \end{ytableau}
\end{align} 
is antisymmetric in the exchange $1\leftrightarrow2$, $6\leftrightarrow7$ and $9\leftrightarrow10$ since $1$, $6$ and $9$ appear in the same column as $2$, $7$ and $10$ respectively.
Consequently, its development on $\gamma_L$ should have for instance a vanishing component on the following SYT of shape $\gamma_L$
\begin{align}
\begin{ytableau}
 1&2 & 7&10\\3 &4&8 \\5&6&9
 \end{ytableau}
\end{align} 
 since this latter is symmetric in the exchange $1\leftrightarrow 2$ and not antisymmetric.
 A systematic way to find the proper expansion with the good coefficients is the following:
 we first create the $f^{\gamma_L}_{\beta_{q_1},\beta_{q_2}}$ SYTs of shape $\gamma_L$ whose $L/2$ first entries are located at the same place as in $S_1$. Here, the $f^{\gamma_L}_{\beta_{q_1},\beta_{q_2}}=f^{[4,3,3]}_{[2,2,1],[2,2,1]}=11$ SYTs are:
 \begin{align}
 \label{basis}
 \begin{ytableau}
 1&3 & 6&7\\2 &4&8 \\5&9&10
 \end{ytableau}
 \,\,\,\begin{ytableau}
 1&3 & 6&8\\2 &4&7 \\5&9&10
 \end{ytableau}\,\,\,
 \,\,\,\begin{ytableau}
 1&3 & 6&7\\2 &4&9 \\5&8&10
 \end{ytableau}\,\,\,
 \begin{ytableau}
 1&3 & 6&8\\2 &4&9 \\5&7&10
 \end{ytableau}
 \,\,\,\begin{ytableau}
 1&3 & 7&8\\2 &4&9 \\5&6&10
 \end{ytableau}\,\,\,
 \begin{ytableau}
 1&3 & 6&9\\2 &4&7 \\5&8&10
 \end{ytableau} \nonumber \\
 \begin{ytableau}
 1&3 & 6&9\\2 &4&8 \\5&7&10
 \end{ytableau}\,\,\,
 \,\,\,\begin{ytableau}
 1&3 & 7&9\\2 &4&8 \\5&6&10
 \end{ytableau}
 \,\,\,\begin{ytableau}
 1&3 & 6&10\\2 &4&7 \\5&8&9
 \end{ytableau}
 \,\,\,\begin{ytableau}
 1&3 & 6&10\\2 &4&8 \\5&7&9
 \end{ytableau}\,\,\,
\begin{ytableau}
 1&3 & 7&10\\2 &4&8 \\5&6&9
 \end{ytableau}
\end{align} 
%$\vert \gamma_L \vert \vert \beta_{q_1},l_1 \otimes \beta_{q_2},l_2\rangle$
Any combination of the last SYTs will have the same internal symmetry properties between the first  $L/2$ entries as in $S_1$.
To find the combination (written like a {\it ket} : $\vert \Phi_{\gamma_L}^{S_1 \otimes S_2 } \rangle $) which will have the same internal symmetry properties between the entries $L/2+1, L/2+2,\cdots,L$ as in $S_2$, one needs now to use the following formula for the quadratic Casimir $C^2$, which holds 
for a general irrep $\beta=[\beta_1,\beta_2,\cdots,\beta_k]$ with $n$ boxes and $k\leq N$ rows:
\begin{align}
\label{formula_casimir}
C^2(\beta)=\sum_{1 \leq i<j \leq n} P_{i,j} =\frac{1}{2}\big\{ \sum_{i}\beta_i^2-\sum_{j}(\beta^T_j)^2\big\}
\end{align}
where the $\beta_i$ are the lengths of the rows, the $\beta^T_j$ are the lengths of the columns (which are also the rows of the transposed shape $\beta^T$),
and the $P_{i,j}$ are the permutations between the indices $i$ and $j$.
Using the relevant indexing for $S_2$, we apply the formula (\ref{formula_casimir}) to the $L/2-1$ {\it successive} shapes which are made of $2$, $3$, $\cdots$, $L/2$ boxes, which contain respectively the set of numbers $\{L, L-1\}$, $\{L,L-1,L-2\}$, and $\{L,L-1,\cdots, L/2+1\}$, and that we write  respectively as $\beta_{q_2}(2),\beta_{q_2}(3),\cdots,\beta_{q_2}(L/2)$. This chain of successive shapes allows one to reconstruct $S_2$.
The application of the formula (\ref{formula_casimir}) to such a chain of shapes  provides a set of $L/2-1$ equations that  $\vert \Phi_{\gamma_L}^{S_1 \otimes S_2 } \rangle $ should satisfy.
So, in our example, one has:
\begin{align}
&\text{\textbullet}P_{9,10}\vert \Phi_{\gamma_L}^{S_1 \otimes S_2 } \rangle =C^2([1,1])\vert \Phi_{\gamma_L}^{S_1 \otimes S_2 } \rangle =-\vert \Phi_{\gamma_L}^{S_1 \otimes S_2 } \rangle,  \nonumber \\
\,\,&\text{since 9 and 10 appear in}\,S_2\,\text{in the subshape}\,\, \ydiagram{1,1}. \label{Chen_equation} \\
\nonumber \\ 
&\text{\textbullet} (P_{9,10}+P_{8,10}+P_{8,9})\vert \Phi_{\gamma_L}^{S_1 \otimes S_2 } \rangle =C^2([1,1,1])\vert \Phi_{\gamma_L}^{S_1 \otimes S_2 } \rangle \nonumber \\ &\hspace{4.75cm} =-3\vert \Phi_{\gamma_L}^{S_1 \otimes S_2 } \rangle,  \,\, \nonumber \\
&\text{since 8, 9 and 10 appear in}\,S_2\,\text{in the subshape}\,\, \ydiagram{1,1,1}. \nonumber \\
\nonumber \\ 
&\text{\textbullet}(P_{9,10}+P_{8,10}+P_{8,9}+P_{7,8}+P_{7,9}+P_{7,10})\vert \Phi_{\gamma_L}^{S_1 \otimes S_2 } \rangle \nonumber \\&=C^2([2,1,1])\vert \Phi_{\gamma_L}^{S_1 \otimes S_2 } \rangle =-2\vert \Phi_{\gamma_L}^{S_1 \otimes S_2 } \rangle , \nonumber \\   \,\,
& \text{since 7, 8, 9 and 10 appear in}\,S_2\,\text{in the subshape}\,\, \ydiagram{2,1,1}. \nonumber \\
 \nonumber \\ 
&\text{\textbullet}(P_{9,10}+P_{8,10}+P_{8,9}+P_{7,8}+P_{7,9}+P_{7,10}+P_{6,7}+P_{6,8}\nonumber \\ &+P_{6,9}+P_{6,10})\vert \Phi_{\gamma_L}^{S_1 \otimes S_2 } \rangle =C^2([2,2,1])\vert \Phi_{\gamma_L}^{S_1 \otimes S_2 } \rangle \nonumber \\ & \hspace{3.7cm} =-2\vert \Phi_{\gamma_L}^{S_1 \otimes S_2 } \rangle,  \,\,\nonumber \\ &\text{since 6, 7,8, 9 and 10 appear in}\,S_2\,\text{in the subshape}\,\, \ydiagram{2,2,1} \nonumber .
\end{align}
As long as $\gamma_L$ appears in $\beta_{q_1}\otimes \beta_{q_2}$ with multiplicity 1 (the number of times that some irrep appears inside a tensor product is called the {\it outer multiplicity}), there is one and only one state $\vert \Phi_{\gamma_L}^{S_1 \otimes S_2 } \rangle $ (up to a phase) which satisfies the last $L/2-1$ equations simultaneously. This is always true in the system under investigation in the current article: antiferromagnetic $SU(N)$ Heisenberg model  with one particle per site in the fundamental irrep, but not always true for the relevant physical problems, as we will discuss in the perspectives. 
 From a practical point of view, one can transform the last problem into the quest of the unique null state of the operator:
 \begin{align}
 \label{null_operator}
 \mathcal{C}_{S_2}=\sum_{q=2}^{L/2}\{\sum_{j=L-q+2}^L  P_{L-q+1,j}&-C^2(\beta_{q_2}(q))\nonumber \\&+C^2(\beta_{q_2}(q-1))\}^2,
 \end{align}
 expressed in the {\it basis} of the $f^{\gamma_L}_{\beta_{q_1},\beta_{q_2}}$  SYTs of shape $\gamma_L$ with  the first $L/2$ entries the same as in $S_1$.
 We use the rules of the orthogonal representation of the symmetric group to write the $f^{\gamma_L}_{\beta_{q_1},\beta_{q_2}} \times  f^{\gamma_L}_{\beta_{q_1},\beta_{q_2}}$ matrix which represents such an operator on such a basis.
For our example, the unique (up to a minus sign) normalized null state of the operator:
\begin{align}
\mathcal{C}_{S_2}=&\{P_{9,10}+1\}^2+\{P_{8,10}+P_{8,9}+2\}^2 \nonumber \\ +&\{P_{7,8}+P_{7,9}+P_{7,10}-1\}^2\nonumber \\ +&\{P_{6,7}+P_{6,8}+P_{6,9}+P_{6,10}\}^2,
\end{align} expressed in the basis shown in (\ref{basis}),
is:
\begin{align}
\label{developpement}
\vert \Phi_{\gamma_L}^{S_1 \otimes S_2 } \rangle =\frac{2}{3} \, &\raisebox{1.4ex}{$\begin{ytableau}
 1&3 & 6&8\\2 &4&9 \\5&7&10
 \end{ytableau}$}\,\,\,+\frac{-\sqrt{2}}{3} \, \raisebox{1.4ex}{$\begin{ytableau}
 1&3 & 7&8\\2 &4&9 \\5&6&10
 \end{ytableau}$}\,\,\,+\frac{\sqrt{2}}{3\sqrt{3}} \, \raisebox{1.4ex}{$\begin{ytableau}
 1&3 & 6&10\\2 &4&8 \\5&7&9
 \end{ytableau}$} \nonumber \\ +\frac{\sqrt{2}}{3\sqrt{3}} \, &\raisebox{1.4ex}{$\begin{ytableau}
 1&3 & 7&9\\2 &4&8 \\5&6&10    \end{ytableau}$}\,\,\,+\frac{-2}{3\sqrt{3}}\, \raisebox{1.4ex}{$\begin{ytableau}
 1&3 & 6&9\\2 &4&8 \\5&7&10
 \end{ytableau}$}\,\,\,+\frac{-1}{3\sqrt{3}} \, \raisebox{1.4ex}{$\begin{ytableau}
 1&3 & 7&10\\2 &4&8 \\5&6&9
 \end{ytableau}$}.
\end{align}

Since there is a freedom in the choice of a sign in the set of real coefficients in Eq. (\ref{developpement}) (one can take $\pm  \vert \Phi_{\gamma_L}^{S_1 \otimes S_2 } \rangle$), one can choose a convention: the coefficient of the first SYT (in the last letter order) in $\vert  \Phi_{\gamma_L}^{S_1 \otimes S_2 } \rangle$ must be taken as positive (equal to $2/3$ in Eq. (\ref{developpement})). This kind of convention is important here because there are consistency relations between the different coefficients (and their different sign) that we are currently calculating. Step 3 is actually motivated by the same reason.

\underline{Step 3}

This stage is a bit technical and is only necessary if $l_2$ is not the highest possible bottom corner in $\beta_{q_2}$. It corresponds to the development around Eq. (4-172) and Eq. (4-173) in \cite{chen}.
The purpose of this stage is to act on $ \vert \Phi_{\gamma_L}^{S_1 \otimes S_2} \rangle $ with the sequence of permutational operations that would move the bottom corner in $S_2$
from its current location (which is the highest possible by construction) to $l_2$, in such a way that one would obtain $S_2'$. The SYT $S_2'$ is defined as the SYT of shape $\beta_{q_2}$, with the number $L/2+1$ located in the bottom corner $l_2$, and with numbers $L/2+2,L/2+3,\cdots,L$ located in the shape $\beta_{q_2}(L/2-1)$ at some positions corresponding to the last SYT of shape $\beta_{q_2}(L/2-1)$ in the last letter order.
The shape $\beta_{q_2}(L/2-1)$ is the shape $\beta_{q_2}$ without the bottom corner located in $l_2$. 
In our example, one has:
\begin{align}
\text{For}\,(\beta_{q_2},l_2)= \raisebox{1.4ex}{$ \ytableaushort{\,\,,\,\,,\times}$}\,\,, S_2=  \raisebox{1.4ex}{$\begin{ytableau}
 10 & 7\\9 &6\\8 \end{ytableau}$}\,\,\,\text{and}\,\,\,S_2'=  \raisebox{1.4ex}{$ \begin{ytableau}
 10 & 8\\9 &7\\6 \end{ytableau}$}.
\end{align}  
In general, the sequence of operations $\mathcal{T}_{S_2}^{S_2'}$ that will transform $S_2$ into $S_2'$ can be written as:
\begin{align}
\mathcal{T}_{S_2}^{S_2'}=\mathcal{T}_{\frac{L}{2}+2}\cdots \mathcal{T}_{k-1}\mathcal{T}_{k},
\end{align}
where $k$ is the largest integer belonging to $\{\frac{L}{2}+2,\frac{L}{2}+3,\cdots,L-1\}$ whose location in $S_2$ is different than its location in $S_2'$,
and where the operator that exchanges numbers $q$ and $q-1$ $\mathcal{T}_{q}$ when acting on a given SYT $S$ can be defined (for $q\geq 2$) as:
\begin{align}
\label{Tau}
\mathcal{T}_{q} S = \frac{P_{q-1,q}-\rho_{q-1,q} \mathcal{I}}{\sqrt{1-\rho_{q-1,q}^2}} S,
\end{align}
where $P_{q-1,q}$ is the permutation between numbers $q$ and $q-1$, $\rho_{q-1,q}$ is the inverse of the axial distance between $q$ and $q-1$ in the SYT S,
and where  $\mathcal{I}$ is the identity. This follows from the expression of the matrix describing permutations of consecutive numbers using SYTs (see appendix).
In our example, one has $k=8$, and:
\begin{align}
\label{T7T8}
\mathcal{T}_{S_2}^{S_2'} S_2 &=\mathcal{T}_{7}\mathcal{T}_{8} S_2 =\mathcal{T}_{7} \mathcal{T}_{8} \, \raisebox{1.4ex}{$ \begin{ytableau}
 10 & 7\\9 &6\\8 \end{ytableau}$} =\mathcal{T}_{7} \frac{P_{7,8}-\rho_{7,8} \mathcal{I}}{\sqrt{1-\rho_{7,8}^2}} \,\raisebox{1.4ex}{$ \begin{ytableau}
 10 & 7\\9 &6\\8 \end{ytableau}$} \nonumber \\ &=\mathcal{T}_{7} \frac{P_{7,8}-\frac{1}{3} \mathcal{I}}{\sqrt{1-\frac{1}{9}}} \,\raisebox{1.4ex}{$ \begin{ytableau}
 10 & 7\\9 &6\\8 \end{ytableau}$}=\frac{P_{6,7}-\frac{1}{2} \mathcal{I}}{\sqrt{1-\frac{1}{4}}} \,\raisebox{1.4ex}{$ \begin{ytableau}
 10 & 8\\9 &6\\7 \end{ytableau}$}=\, \raisebox{1.4ex}{$ \begin{ytableau}
 10 & 8\\9 &7\\6 \end{ytableau}$}=S_2'.
\end{align}
We then act with $\mathcal{T}_{S_2}^{S_2'}$ on $\vert \Phi_{\gamma_L}^{S_1 \otimes S_2 } \rangle $ to obtain $\vert \Phi_{\gamma_L}^{S_1 \otimes S_2'} \rangle $; here, using Eq. (\ref{T7T8}) , Eq. (\ref{developpement}),  and the rules for the orthogonal representation of the symmetric group, we get:
\begin{align}
\label{final_dev_1}
\mathcal{T}_{S_2}^{S_2'} \vert \Phi_{\gamma_L}^{S_1 \otimes S_2} \rangle&= \frac{P_{6,7}-\frac{1}{2} \mathcal{I}}{\sqrt{1-\frac{1}{4}}}\times \frac{P_{7,8}-\frac{1}{3} \mathcal{I}}{\sqrt{1-\frac{1}{9}}} \vert \Phi_{\gamma_L}^{S_1 \otimes S_2} \rangle  \\ &= \vert \Phi_{\gamma_L}^{S_1 \otimes S_2'}  \rangle = \frac{1}{4} \, \raisebox{1.4ex}{$\begin{ytableau}
 1&3 & 6&10\\2 &4&8 \\5&7&9
 \end{ytableau}$}\,\,\,+\frac{-\sqrt{3}}{4}\, \raisebox{1.4ex}{$\begin{ytableau}
 1&3 & 6&10\\2 &4&7 \\5&8&9
 \end{ytableau}$}\,\,\,\nonumber \\ &\hspace{1.35cm}+\frac{\sqrt{3}}{2\sqrt{2}} \, \raisebox{1.4ex}{$\begin{ytableau}
 1&3 &6&9\\2 &4&7 \\5&8&10
 \end{ytableau}$}\,\,\,\,+\frac{-1}{2\sqrt{2}} \, \raisebox{1.4ex}{$\begin{ytableau}
 1&3 & 6&9\\2 &4&8 \\5&7&10
 \end{ytableau}$}\,\,\,\nonumber \\ &\hspace{1.35cm}+\frac{-\sqrt{3}}{4\sqrt{2}} \, \raisebox{1.4ex}{$\begin{ytableau}
 1&3 & 6&8\\2 &4&9 \\5&7&10    \end{ytableau}$}\,\,\,+\frac{\sqrt{5}}{4\sqrt{2}} \, \raisebox{1.4ex}{$\begin{ytableau}
 1&3 & 6&7\\2 &4&9 \\5&8&10 \nonumber
 \end{ytableau}$}.
\end{align}
The coefficients appearing in the right hand side of Eqs. (\ref{developpement}) and (\ref{final_dev_1}) are the $S_L \subset S_{\frac{L}{2}}\otimes S_{\frac{L}{2}}$  {\it subduction coefficients} \cite{chen},
(where $S_n$ stands for the symmetric group of n elements)
\\
Now, one can wonder why we have not taken directly $S_2'$ at then end of Step 1 in order to start Step 2 with the expansion of the product $S_1\otimes S_2'$ and not $S_1\otimes S_2$. If we did so, we would obtain in general the same combination as $\vert \Phi_{\gamma_L}^{S_1 \otimes S_2'} \rangle$ up to a sign.
To ensure some consistency relations between  subduction coefficients, as first pointed out by Chen \cite{chen}, one needs to start doing calculation on $S_2$, the {\it last} SYT of shape $\beta_{q_2}$. Let us stress that these signs are not just a matter of convention. It is absolutely necessary to keep track of them to get the matrix right.

\underline{Step 4 \& 5\& 6}

We perform the steps 1 \& 2 and 3 but with $(\beta_{q_3},l_3)$ instead of $(\beta_{q_1},l_1)$ and $(\beta_{q_4},l_4)$ instead of $(\beta_{q_2},l_2)$.
In our example, we have:
\begin{align}
S_3=\begin{ytableau}
 1& 3&5\\2 &4
 \end{ytableau}\,\,\text{and}\,\,S_4=S_4'=\begin{ytableau}
 10& 8&6\\9 &7
 \end{ytableau},
\end{align} 
and
\begin{align}
\label{final_dev_2}
\vert \Phi_{\gamma_L}^{S_3 \otimes S_4'} \rangle= \frac{1}{4} \, &\raisebox{1.4ex}{$\begin{ytableau}
 1&3 & 5&10\\2 &4&8 \\6&7&9
 \end{ytableau}$}\,\,\,+\frac{-\sqrt{3}}{4}\, \raisebox{1.4ex}{$\begin{ytableau}
 1&3 & 5&10\\2 &4&7 \\6&8&9
 \end{ytableau}$}\,\,\,+\frac{\sqrt{3}}{2\sqrt{2}} \, \raisebox{1.4ex}{$\begin{ytableau}
 1&3 &5&9\\2 &4&7 \\6&8&10
 \end{ytableau}$}\nonumber \\ +\frac{-1}{2\sqrt{2}} \, &\raisebox{1.4ex}{$\begin{ytableau}
 1&3 & 5&9\\2 &4&8 \\6&7&10
 \end{ytableau}$}\,\,\,+\frac{-\sqrt{3}}{4\sqrt{2}} \, \raisebox{1.4ex}{$\begin{ytableau}
 1&3 & 5&8\\2 &4&9 \\6&7&10    \end{ytableau}$}\,\,\,+\frac{\sqrt{5}}{4\sqrt{2}} \, \raisebox{1.4ex}{$\begin{ytableau}
 1&3 & 5&7\\2 &4&9 \\6&8&10
 \end{ytableau}$}.
\end{align}
In our example, the coefficients in $ \vert \Phi_{\gamma_L}^{S_1 \otimes S_2'} \rangle $ (Eq. (\ref{final_dev_1})) are the same as in $\vert  \Phi_{\gamma_L}^{S_3 \otimes S_4'} \rangle $ (Eq. (\ref{final_dev_2})): it is of course not always true.

\underline{Step 7}

We finally use the Young's rules (cf section \ref{young_rep}) to calculate:
\begin{align}
&\langle \beta_{q_3},l_3  \vert \otimes\langle \beta_{q_4},l_4 \vert  \mathcal{P}_{\gamma_L} \vert \beta_{q_1},l_1\rangle \otimes \vert \beta_{q_2},l_2\rangle \nonumber \\ &=\langle \Phi_{\gamma_L}^{S_3 \otimes S_4'} \vert P_{\frac{L}{2},\frac{L}{2}+1} \vert \Phi_{\gamma_L}^{S_1 \otimes S_2'} \rangle.
\end{align}
In our example, we obtain:
\begin{align}
\langle \Phi_{\gamma_L}^{S_3 \otimes S_4'} \vert P_{\frac{L}{2},\frac{L}{2}+1} \vert \Phi_{\gamma_L}^{S_1 \otimes S_2'} \rangle=\frac{\sqrt{15}}{4}.
\end{align}

 \subsubsection{Improvement of Chen's method: taking a "shortcut" in the chain of subshapes}
 
The method developed in the last section is already quite powerful and would enable us to have accurate DMRG simulations for $N=3$ or $N=4$.
However, when $N$ increases, it becomes limited due to the high value of $f^{\gamma_L}_{\beta_{q_1},\beta_{q_2}}$, which determines the size of the matrix to obtain the unique null state of Step 2.
We found a way to improve a lot Step 2 by taking advantage of the {\it lego-like} structure of $SU(N)$ irreps and SYTs.
In fact, every shape or SYT with $n$ boxes can either be seen as living in the tensor product of $n$ fundamental irreps, or as living in the tensor product of non fundamental irreps, such as one-column fully antisymmetric irreps for instance. 
Based on this principle, our method below is presented: it will lead to an important reduction of the size of the  matrix to obtain the null state of in Step 2.
For instance, for SU(6), and for $\beta_{q_1}=[4,3,3,2,2,2]$, $\beta_{q_2}=[3,3,3,3,3,1]$, there are $f^{\gamma_L}_{\beta_{q_1},\beta_{q_2}}=1081080$ SYTs of shape $\gamma_L=[6,6,5,5,5,5]$, involving the necessity to find the null state of a matrix of size $1081080\times1081080$, while the matrix that we need to find the null state of in the current section has dimension $2 \times 2$.
In this new example, one has (for $l_1=6$):
\begin{align}
\label{S_1S_2antisym1}
S_1=  \raisebox{1.4ex}{$ \scalebox{1.3}
{\begin{ytableau}
 1 & 7 &12 &15\\2 &8&13\\3&9&14 \\4&10\\5&11\\6&16 \end{ytableau}}$}\,\,\,\text{and}\,\,\,S_2=  \raisebox{1.4ex}{$\scalebox{1.3}
{ \begin{ytableau}
\,\,\,32 \,\,\,& \,\,\, 26 \,\,\,& \,\,\,21\,\,\,\\ \,\,\,31\,\,\, &\,\,\, 25\,\,\, &\,\,\, 20 \,\,\, \\ \,\,\, 30\,\,\, & \,\,\,24\,\,\, & \,\,\,19\,\,\, \\ \,\,\,29\,\,\, & \,\,\,23\,\,\, &\,\,\, 18\,\,\, \\ \,\,\,28\,\,\, & \,\,\, 22\,\,\, &\,\,\, 17\,\,\, \\ \,\,\,27 \,\,\, \end{ytableau}}$}.
\end{align}
In Chen's method, $S_2$ would be characterized by the chain of successive subshapes containing the set of numbers $\{L,L-1\}$, $\{L,L-1,L-2\}$,  $\{L,L-1,L-3,\}$,  $\{L,L-1,L-2,\cdots,L/2+1\}$.
In the current example, it would be:
\begin{align}
&\ydiagram{1} \rightarrow \ydiagram{1,1} \rightarrow \ydiagram{1,1,1} \rightarrow \ydiagram{1,1,1,1} \rightarrow \ydiagram{1,1,1,1,1} \rightarrow \ydiagram{1,1,1,1,1,1} \rightarrow  \ydiagram{2,1,1,1,1,1} \rightarrow \ydiagram{2,2,1,1,1,1} \rightarrow \ydiagram{2,2,2,1,1,1} \nonumber \\ &\rightarrow \ydiagram{2,2,2,2,1,1}  \rightarrow  \ydiagram{2,2,2,2,2,1} \rightarrow  \ydiagram{3,2,2,2,2,1} \rightarrow  \ydiagram{3,3,2,2,2,1} \rightarrow  \ydiagram{3,3,3,2,2,1} \rightarrow  \ydiagram{3,3,3,3,2,1}\nonumber \\ & \rightarrow  \ydiagram{3,3,3,3,3,1}.
\end{align}
Then, from this chain, we would build the operator  $\mathcal{C}_{S_2}$ defined in Eq. (\ref{null_operator}). We would then look for  the unique combination of SYTs of shape $\gamma_L$, null on $ \mathcal{C}_{S_2}$ and built from  the tensor product of $S_1 \in \beta_{q_1}$  and 
\begin{align}
S_2 \in \beta_{q_2} \in \ydiagram{1}^{\otimes \frac{L}{2}}.
\end{align}
In the present section, we characterize $S_2$ by the chain of subshapes containing the set of numbers $\{L,L-1,L-2,\cdots,L+1-\beta_{q_2}^{T,1}\}$,  $\{L,L-1,L-2,\cdots,L+1-\beta_{q_2}^{T,1},L-\beta_{q_2}^{T,1}, \cdots,L-\beta_{q_2}^{T,1}-\beta_{q_2}^{T,2}+1,\}$,$\cdots$,$\{L,L-1,L-2,\cdots,L/2+1\}$,
where the numbers $\beta_{q_2}^{T,j}$ stand for the length of the $j^{th}$ column in $\beta_{q_2}$, for $j=1,\cdots,n^c_{\beta_{q_2}}$, with $n^c_{\beta_{q_2}}$ the number of columns in the shape $\beta_{q_2}$.
In the current example, it is the chain:
\begin{align}
\ydiagram{1,1,1,1,1,1} \rightarrow \ydiagram{2,2,2,2,2,1} \rightarrow  \ydiagram{3,3,3,3,3,1}. \label{chain_new}
\end{align}

We will then look for all the states which live in the irrep represented by the shape $\gamma_L$, and which are built from  the tensor product of the state $S_1 \in \beta_{q_1}$  and the states 
\begin{align}
S_2 \in \beta_{q_2} \in 
\begin{ytableau} \, \\ \, \\ \none[\scalebox{0.5}{\vdots}] \\ \, \\ \, \end{ytableau} \,  \otimes \begin{ytableau} \, \\ \, \\ \none[\scalebox{0.5}{\vdots}]  \\ \, \end{ytableau}\otimes  \scalebox{0.5}{...}\,\otimes \begin{ytableau} \, \\ \none[\scalebox{0.5}{\vdots}]  \\ \, \end{ytableau},
\end{align}
which means that $\beta_{q_2}$ is now seen as an irrep produced by the tensor product  of the one-column irreps of  length $\beta_{q_2}^{T,k}$ from $k=1$ to $k=n^c_{\beta_{q_2}}$.
These states are superpositions of SYTs of shape $\gamma_L$, which all have their first $L/2$ entries located as in $S_1$, and their entries $\{L/2+1,L/2+2,\cdots,L\}$ which satisfy some  internal constraints corresponding to the product of $n^c_{\beta_{q_2}}$ fully antisymmetric irreps of length $\beta_{q_2}^{T,k}$ from $k=1$ to $k=n^c_{\beta_{q_2}}$. Such a set of internal constraints for the case of antisymmetric irrep has firstly been pointed out in \cite{PRBNataf2016}.Let's call $\text{Col}^{S_2}_k$, the set of numbers located in the $k^{th}$ column of $S_2$ (for $k=1,\cdots,n^c_{\beta_{q_2}}$). The internal constraints here are the following: every SYT should be such that the set of decreasing numbers belonging to $\text{Col}^{S_2}_k$ must appear in the shape $\gamma_L$ in  stricly ascending rows from bottom to top, for $k=1, \cdots, n^c_{\beta_{q_2}}$. 
The number of such SYTs will be named $\tilde{f}^{\gamma_L}_{\beta_{q_1},\beta_{q_2}}$, and is often much smaller than $f^{\gamma_L}_{\beta_{q_1},\beta_{q_2}}$.
In our example, $\tilde{f}^{\gamma_L}_{\beta_{q_1},\beta_{q_2}}=2$, and the $2$ SYTs which satisfy all the constraints are:
\begin{align}
\label{S_1S_2antisym2}
&\tilde{S}^{\gamma_L,1}_{\beta_{q_1},\beta_{q_2}}=\scalebox{1.3}
{\begin{ytableau}
 1 & 7 &12 &15 & 22 & 27\\2 &8&13 & 17 & 23 & 28 \\3&9&14 & 18 & 29 \\4&10 & 19 & 24 & 30 \\ 5 &11&20&25&31\\6&16&21&26&32 \end{ytableau}}, \nonumber \\ 
&\tilde{S}^{\gamma_L,2}_{\beta_{q_1},\beta_{q_2}}=\scalebox{1.3}
{ \begin{ytableau}
 1 & 7 &12 &15 & 17 & 27\\2 &8&13 & 18 & 22 & 28 \\3&9&14 & 23 & 29 \\4&10 & 19 & 24 & 30 \\ 5 &11&20&25&31\\6&16&21&26&32
  \end{ytableau}},
 \end{align}
where for instance, one can chek that the set of numbers $\text{Col}^{S_2}_1=\{32,31,30,\cdots,27\}$ which appear in the first column of $S_2$ in Eq. (\ref{S_1S_2antisym1}),
are located in the two SYTs of Eq. (\ref{S_1S_2antisym2}) in strictly ascending rows from $32$ to $27$.
Using the words and concepts developped in \cite{PRBNataf2016,Wan2017}, these two SYT must be seen as {\it representatives} of {\it equivalence classes}.
The SYT $\tilde{S}^{\gamma_L,j}_{\beta_{q_1},\beta_{q_2}}$ (for some $j$ such that $1\leq j \leq  \tilde{f}^{\gamma_L}_{\beta_{q_1},\beta_{q_2}}$) represents all the SYTs of same shape $\gamma_L$, with entries from $1$ to $L/2$ at the same location as in $S_1$ and with entries from $L/2+1$ to $L$ being at the same location as in $\tilde{S}^{\gamma_L,j}_{\beta_{q_1},\beta_{q_2}}$ up to some permutation among the numbers along each column in $S_2$.
For instance, the SYTs:
\begin{align}
\scalebox{1.3}
{\begin{ytableau}
 1 & 7 &12 &15 & 22 & 27\\2 &8&13 & 17 & 24 & 29 \\3&9&14 & 18 & 28 \\4&10 & 19 & 23 & 30 \\ 5 &11&20&25&31\\6&16&21&26&32 \end{ytableau}}
 \,\,\,\text{,}\,\,\,
\scalebox{1.3}
{ \begin{ytableau}
 1 & 7 &12 &15 & 22 & 29\\2 &8&13 & 17 & 23 & 30 \\3&9&14 & 18 & 27 \\4&10 & 19 & 24 & 28 \\ 5 &11&20&25&31\\6&16&21&26&32 
  \end{ytableau}}
   \,\,\,\text{and}\,\,\,
\scalebox{1.3}
{ \begin{ytableau}
 1 & 7 &12 &15 & 22 & 29\\2 &8&13 & 19 & 23 & 30 \\3&9&14 & 20 & 27 \\4&10 & 17 & 24 & 28 \\ 5 &11&18&25&31\\6&16&21&26&32 
  \end{ytableau}}
 \end{align}
are three SYTs (among many others) which belong to the equivalence class represented by the SYT $\tilde{S}^{\gamma_L,1}_{\beta_{q_1},\beta_{q_2}}$ appearing in Eq. (\ref{S_1S_2antisym2}).
From each SYT  $\tilde{S}^{\gamma_L,j}_{\beta_{q_1},\beta_{q_2}}$ (for $j$ such that $1\leq j \leq  \tilde{f}^{\gamma_L}_{\beta_{q_1},\beta_{q_2}}$), we then construct a {\it state} $\vert  \tilde{\phi} ^{\gamma_L,j}_{\beta_{q_1},\beta_{q_2}} \rangle$, which is a  linear combination of SYTs belonging to the same class, obtained by using some Projection operator \cite{PRBNataf2016,Wan2017} which ensures the property of antisymmetry among the numbers located in each column of $S_2$: $\vert \tilde{\phi} ^{\gamma_L,j}_{\beta_{q_1},\beta_{q_2}} \rangle= \mathcal{N}^{-1} Proj \tilde{S}^{\gamma_L,j}_{\beta_{q_1},\beta_{q_2}}$.

We then define  a new operator $\tilde{\mathcal{C}}_{S_2}$ as:
\begin{align}
 \tilde{\mathcal{C}}_{S_2}=\sum_{k=1}^{n^c_{\beta_{q_2}}-1}\{P_{n_{k}^{S_2},n_{k}^{S_2}+1}-\frac{1}{\beta_{q_2}^{T,k}}\}^2,
\end{align}
where the number $n_k^{S_2}=L+1-\sum_{r=1}^{k}\beta_{q_2}^{T,k}$ (for $k=1,\cdots,n^c_{\beta_{q_2}}$) is the number located in the last row of the $k^{th}$ column in $S_2$.
In our example,
$ \tilde{\mathcal{C}}_{S_2}$ would be :
\begin{align}
 \tilde{\mathcal{C}}_{S_2}=\{P_{26,27}-\frac{1}{6}\}^2+\{P_{21,22}-\frac{1}{5}\}^2.
\end{align}
The state $\vert \Phi _{\gamma_L}^{S_1\otimes S_2} \rangle$ (which was defined in the previous section) can be obtained by looking for the (unique) null state of $\tilde{\mathcal{C}}_{S_2}$ expressed in the {\it basis} of states
$\vert  \tilde{\phi} ^{\gamma_L,j}_{\beta_{q_1},\beta_{q_2}} \rangle$ . 
To prove such a claim, we use the formula (\ref{formula_casimir}) applied on the subshape made of the $k^{th}$ column in $S_2$ (of length $\beta_{q_2}^{T,k}$) and the highest box of the  $(k+1)^{th}$ column of $S_2$ (for $k=1,\cdots,n^c_{\beta_{q_2}}-1$), which is filled by the set of consecutive numbers $\text{Col}^{S_2}_k \cup \{n_k^{S_2}-1 \} = \{n_k^{S_2}+\beta_{q_2}^{T,k}-1,n_k^{S_2}+\beta_{q_2}^{T,k}-2,\cdots,n_k^{S_2}, n_k^{S_2}-1\}$:
\begin{align}
&\sum_{ \substack{i,j\in \text{Col}^{S_2}_k \cup \{n_k^{S_2}-1 \}  \\ i<j}} \hspace{-.7cm}P_{i,j} \vert \Phi _{\gamma_L}^{S_1\otimes S_2} \rangle=\Big{\{}1+ \beta_{q_2}^{T,k}\frac{1-\beta_{q_2}^{T,k}}{2} \Big{\}} \vert \Phi _{\gamma_L}^{S_1\otimes S_2} \rangle \nonumber \\ & \text{since} \,\,\text{Col}^{S_2}_k \cup \{n_k^{S_2}-1 \}  \,\text{appear in}\, S_2\, \text{in the subshape:} \nonumber \\ &\raisebox{1 ex}{$\mathsmaller{\mathsmaller{\mathsmaller{\beta_{q_2}^{T,k}}}}$} \raisebox{2.5ex}{$\raisebox{-3.5ex}{$\scalebox{3}{\{}$} \begin{ytableau} \,& \\ \, \\ \none[\scalebox{0.5}{\vdots}]  \\ \, \end{ytableau}.$}
\end{align}
Now, since:
\begin{align}
\sum_{ \substack{i,j\in \text{Col}^{S_2}_k \cup \{n_k^{S_2}-1 \}  \\ i<j}} P_{i,j}=\sum_{ \substack{i\in \text{Col}^{S_2}_k \\ j=n_k^{S_2}-1}} P_{i,j}+\sum_{ \substack{i,j\in \text{Col}^{S_2}_k \\ i<j}} P_{i,j},
\end{align}
%and since every basis state $\vert  \tilde{\phi} ^{\gamma_L,m}_{\beta_{q_1},\beta_{q_2}} \rangle$ (for $m$ such that $1\leq m \leq  \tilde{f}^{\gamma_L}_{\beta_{q_1},\beta_{q_2}}$)
%is fully antisymmetric (by construction) in the exchange $i \leftrightarrow j$, where $i,j\in \text{Col}^{S_2}_k$, one should have:
and since $\vert \Phi _{\gamma_L}^{S_1\otimes S_2} \rangle$ should be fully antisymmetric in the exchange $i \leftrightarrow j$, where $i,j\in \text{Col}^{S_2}_k$, one should have:

\begin{align}
\sum_{ \substack{i\in \text{Col}^{S_2}_k \\ j=n_k^{S_2}-1}} P_{i,j} \vert \Phi _{\gamma_L}^{S_1\otimes S_2} \rangle= \vert \Phi _{\gamma_L}^{S_1\otimes S_2}\rangle.
\end{align}
Secondly, if we also use the  antisymmetry for the numbers inside $\text{Col}^{S_2}_{k+1}$, and if we write $P_{i,j}=P_{j,n_k^{S_2}-1} P_{i,n_k^{S_2}-1} P_{j,n_k^{S_2}-1}$ %=P_{i,n_k^{S_2}} P_{j,n_k^{S_2}-1} P_{n_k^{S_2},n_k^{S_2}-1} P_{i,n_k^{S_2}} P_{j,n_k^{S_2}-1} $ 
%for every basis state $\vert  \tilde{\phi} ^{\gamma_L,m}_{\beta_{q_1},\beta_{q_2}} \rangle$ (for $m$ such that $1\leq m \leq  \tilde{f}^{\gamma_L}_{\beta_{q_1},\beta_{q_2}}$) 
for $i\in \text{Col}^{S_2}_k$ and for $j\in \text{Col}^{S_2}_{k+1}$, we obtain, $\forall k=1, \cdots, n^c_{\beta_{q_2}}-1$:
\begin{align}
H^{\text{Col},S_2}_{k,k+1}\vert \Phi _{\gamma_L}^{S_1\otimes S_2}\rangle=\sum_{ \substack{i\in \text{Col}^{S_2}_k \\ j\in \text{Col}^{S_2}_{k+1}}} P_{i,j}\vert \Phi _{\gamma_L}^{S_1\otimes S_2} \rangle= \beta_{q_2}^{T,k+1} \vert \Phi _{\gamma_L}^{S_1\otimes S_2}\rangle,\label{equation_new_method}
\end{align}
where we have also defined $H^{\text{Col},S_2}_{k,k+1}$ as the sum of $ \beta_{q_2}^{T,k+1}\times \beta_{q_2}^{T,k}$ permutations $P_{i,j}$ that interexchange numbers $i\in\text{Col}^{S_2}_{k}$ with numbers  $j\in\text{Col}^{S_2}_{k+1}$.
This set of $n^c_{\beta_{q_2}}-1$ equations (Eq. (\ref{equation_new_method})) now replaces the set of $L/2-1$ equations appearing in Eq. (\ref{Chen_equation}) for the case treated using Chen's method. It should be solved in the basis of states $\{ \vert  \tilde{\phi} ^{\gamma_L,m}_{\beta_{q_1},\beta_{q_2}} \rangle, m = 1,\cdots,  \tilde{f}^{\gamma_L}_{\beta_{q_1},\beta_{q_2}} \}$ which are antisymmetric on each set of numbers $\text{Col}^{S_2}_{k}$ (for $k=1, \cdots, n^c_{\beta_{q_2}}$) by construction. This set of equations uniquely determines $\vert \Phi _{\gamma_L}^{S_1\otimes S_2} \rangle$ because the eigenvalues of the operator $H^{\text{Col},S_2}_{k,k+1}$ on such a basis are $\{\beta_{q_2}^{T,k+1},\beta_{q_2}^{T,k+1}-1,\beta_{q_2}^{T,k+1}-2,\cdots\}$  and are associated to the following tensor product:
\begin{widetext}
\begin{align}
&\raisebox{-1ex}{$\mathsmaller{\mathsmaller{\mathsmaller{\beta_{q_2}^{T,k}}}}$} \raisebox{2.5ex}{ $\raisebox{-6ex}{$\scalebox{4}{\{}$} \begin{ytableau} \, \\ \, \\ \, \\ \, \\ \none[\scalebox{0.5}{\vdots}]  \\ \, \end{ytableau}$} \,\,\raisebox{1ex}{$\otimes$}  \,\raisebox{2.5ex}{$\begin{ytableau} \, \\  \, \\ \none[\scalebox{0.5}{\vdots}]  \\ \, \end{ytableau}$} \raisebox{-1ex}{$\scalebox{3}{\}}$}\mathsmaller{\mathsmaller{\mathsmaller{\beta_{q_2}^{T,k+1}}}} =\raisebox{-1ex}{$\mathsmaller{\mathsmaller{\mathsmaller{\beta_{q_2}^{T,k}}}}$}\raisebox{2.5ex}{$\raisebox{-6ex}{$\scalebox{4}{\{}$} \begin{ytableau} \,&\, \\ \,&\, \\ \,&\none[\scalebox{0.5}{\vdots}]  \\ \,&\\ \none[\scalebox{0.5}{\vdots}]  \\ \, \end{ytableau}$}  \raisebox{-1ex}{$\scalebox{3}{\}}$}\mathsmaller{\mathsmaller{\mathsmaller{\beta_{q_2}^{T,k+1}}}} \oplus \hspace{-.2cm}   \raisebox{-2ex}{$\mathsmaller{\mathsmaller{\mathsmaller{\beta_{q_2}^{T,k}+1}}}$}\raisebox{2.5ex}{$\raisebox{-7ex}{$\scalebox{4.5}{\{}$} \begin{ytableau} \,&\, \\ \,&\none[\scalebox{0.5}{\vdots}]\, \\ &  \\ \,\\ \none[\scalebox{0.5}{\vdots}]  \\ \,\\ \, \end{ytableau}$}  \raisebox{-0.0ex}{$\scalebox{2.5}{\}}$}\raisebox{1.5ex}{$\mathsmaller{\mathsmaller{\mathsmaller{\beta_{q_2}^{T,k}-1}}}$} \oplus \cdots\oplus \hspace{-2cm}\raisebox{-3ex} {$\mathsmaller{\mathsmaller{\mathsmaller{r=Max(\beta_{q_2}^{T,k}+\beta_{q_2}^{T,k+1},N)}}} $}\raisebox{2.5ex}{$\raisebox{-8ex}{$\scalebox{5}{\{}$} \begin{ytableau} \,&\, \\ \,&\none[\scalebox{0.5}{\vdots}] \\ \,  \\ \,\\ \none[\scalebox{0.5}{\vdots}]  \\ \,\\ \,\\ \, \end{ytableau}$}  \raisebox{1ex}{$\scalebox{2}{\}}$}\raisebox{2ex}{$\mathsmaller{\mathsmaller{\mathsmaller{\beta_{q_2}^{T,k+1}-(r-\beta_{q_2}^{T,k}) }}}$}\\
&H^{\text{Col},S_2}_{k,k+1\hspace{2.5cm}}\rightarrow \hspace{1cm}  \beta_{q_2}^{T,k+1} \hspace{3cm}\beta_{q_2}^{T,k+1}-1\hspace{1cm} .....\hspace{1.5cm} \beta_{q_2}^{T,k+1}-(r-\beta_{q_2}^{T,k})  \nonumber
\label{tensor_product_antisymm_3}
\end{align}
\end{widetext}
where the size of the columns are put next to the  curly brackets, and where the value taken by $H^{\text{Col},S_2}_{k,k+1}$ on each $SU(N)$ irrep on the right hand side appears below the corresponding $SU(N)$ irrep: it is nothing but the lengths of the second columns of these irreps. To calculate these quantities,
we have used the same steps as those used to derive (Eq. (\ref{equation_new_method})). Thus, the set of equations (Eq. (\ref{equation_new_method})) is enough to impose the chain of subshapes shown in Eq. (\ref{chain_new}) for the example treated here.
Finally, the null state of :
\begin{align}
 \sum_{k=1}^{n^c_{\beta_{q_2}}-1}\{ H^{\text{Col},S_2}_{k,k+1}- \beta_{q_2}^{T,k+1} \}^2
\end{align} on the basis  $\{ \vert  \tilde{\phi} ^{\gamma_L,m}_{\beta_{q_1},\beta_{q_2}} \rangle, m = 1,\cdots,  \tilde{f}^{\gamma_L}_{\beta_{q_1},\beta_{q_2}} \} $ is also the null state of $\tilde{\mathcal{C}}_{S_2}$ on such a basis since we have:

\begin{align}
&\langle  \tilde{\phi}^{\gamma_L,m'}_{\beta_{q_1},\beta_{q_2}} \vert H^{\text{Col},S_2}_{k,k+1} \vert  \tilde{\phi} ^{\gamma_L,m}_{\beta_{q_1},\beta_{q_2}} \rangle \nonumber \\ &=  \beta_{q_2}^{T,k+1}  \beta_{q_2}^{T,k}\langle  \tilde{\phi}^{\gamma_L,m'}_{\beta_{q_1},\beta_{q_2}}   \vert  P_{n_{k}^{S_2},n_{k}^{S_2}+1} \vert \tilde{\phi} ^{\gamma_L,m}_{\beta_{q_1},\beta_{q_2}} \rangle,
\end{align}
thanks also to the antisymmetry in $\text{Col}^{S_2}_k$ and $\text{Col}^{S_2}_{k+1}$. 
Note that if $\tilde{f}^{\gamma_L}_{\beta_{q_1},\beta_{q_2}}=1$, one does not need to create  $\tilde{\mathcal{C}}_{S_2}$ because one will directly have $\vert \Phi _{\gamma_L}^{S_1\otimes S_2} \rangle=\vert  \tilde{\phi} ^{\gamma_L,1}_{\beta_{q_1},\beta_{q_2}} \rangle$. 
Such a case for instance occurs when the shape $\beta_{q_2}$ has only one column.
Finally, from the coefficients of the null state of $\tilde{\mathcal{C}}_{S_2}$ and from the developments of the states $\{ \vert  \tilde{\phi} ^{\gamma_L,m}_{\beta_{q_1},\beta_{q_2}} \rangle, m = 1,\cdots,  \tilde{f}^{\gamma_L}_{\beta_{q_1},\beta_{q_2}} \}$, one recovers the subduction coefficients the way they would appear at the end of step 2.
But the computation is much faster.

%\begin{align}
%&\text{\textbullet}
%\end{align}

 %state which vanishes under the action of  $\tilde{\mathcal{C}}_{S_2}$

\section{Bethe Ansatz Equations for the Heisenberg $SU(N)$ chain with open boundary conditions}
\label{exact}
In order to benchmark our DMRG code, we need to compare our  numerical results to some exact results.
The set of Bethe Ansatz Equations for a $SU(N)$ Heisenberg chain of $L$ sites with Periodic Boundary Conditions (PBC) appears explicitely in Eq. (5.2 b) of [\cite{johannesson1986}].
Such a set of equations has for instance been used by Alcaraz and Martins in [\cite{Alcaraz_1989}] to obtain, among other results, the finite size ring ground state energies for the SU(3) and SU(4) Heisenberg chain with the fundamental irrep at each site.\\
Many interesting studies on the integrable $SU(N)$ Heisenberg chains with different boundary conditions (and different irreps at each site) have been done in the past\cite{doikou_1998,frappat2004,frappat2007}; however we have not been able to find in the literature the Bethe ansatz equations (BAE) for open boundary conditions (OBC) and the fundamental irrep in an explicit form that would be as simple to use as that of Eq. (5.2 b) in  [\cite{johannesson1986}] (which is valid only for PBC), with the noticeable exception of a recent  study of the $SU(3)$ case [\cite{Wen_SU3_exact}].
So, as a first step, we write below the set of BAE for an $L$ sites $SU(N)$ Heisenberg chain with OBC with the fundamental irrep at each site for general $N$:
\begin{widetext}
\begin{align}
\label{BAE}
\arctan{\big{(}2x_k^{(r)}\big{)}}&+\sum_{\alpha=r\pm1}\sum_{j=1}^{M^{(\alpha)}}\big{\{}\arctan{\big{(}2[x_k^{(r)}-x_j^{(\alpha)}]\big{)}}+\arctan{\big{(}2[x_k^{(r)}+x_j^{(\alpha)}]\big{)}}\big{\}}
\nonumber \\&-\sum_{j=1}^{M^{(r)}}\big{\{}\arctan{\big{(}x_k^{(r)}-x_j^{(r)}\big{)}}+\arctan{\big{(}x_k^{(r)}+x_j^{(r)}\big{)}}\big{\}}=\pi I_k^{(r)},\,\,\,\,\, k=1,\cdots,M^{(r)},
\end{align}
\end{widetext}
for $r=1,\cdots,N-1$.
For the ground state, the quantum numbers $M^{(r)}$ and $I_k^{(r)}$ are integers given by:
\begin{align}
M^{(r)}&=\text{Floor}\big{[} L(1-\frac{r}{N}) \big{]},\\
I_k^{(r)}&=k, \,\,\,\text{for}\,\,\,k=1,2,\cdots,M^{(r)},
\end{align}
for $r=1,\cdots,N-1$, while $M^{(N)}=0$ and $M^{(0)}=L$. Finally, one should take $x_j^{(0)}=0$ $\forall j=1,\cdots,L=M^{(0)}$.
So, for instance, for $L=12$ and $SU(3)$, $M^{(1)}=8$, $M^{(2)}=4$, so that there are $12$ unknown {\it Bethe roots}  $\{x_k^{(r)}\}$,  $(r=1,2)$ that one can obtain numerically by solving the system of equations Eq. (\ref{BAE}).

Once the Bethe roots $x_l^{(1)}$ for $l=1..M^{(1)}$ have been determined, the ground state energy of the open chain of $L$ sites can be obtained as:
\begin{align}
E(L)=-\sum_{l=1}^{M^{(1)}}\frac{1}{\frac{1}{4}+(x_l^{(1)})^2}+(L-1),
\end{align}
In Table \ref{table_EGS_exact}, we list for some values of $L$ and $N$ the ground state energy per site $E(L)/L$. These values will be used in the next section to benchmak our DMRG code.

%\begin{figure} 
%\centerline{\includegraphics[width=\linewidth]{entanglement_SU3_1part_240}}
%\caption{Entanglement Entropy for one particle per site for $L=240$ sites. In the inset of the right figure, central charges as a function of the lost weight for $m=500$, $m=1000$ and $m=2000$ states kept.}
%\end{figure}

\begin{table*}[ht]
        \centering
        \begin{tabular}{|c|c|c|c|c|c|c|c|}
                \hline
                L&SU(3)&SU(4)&SU(5)&SU(6)&SU(8) \\
                \hline
                60&-0.6998571791373&-0.8188090258187&-0.8765610853081&-0.9088545631758&-0.941247635.......\footnote{There are less digits because the convergence of the numerical solver is not as good when L is not a multiple of N.} \\
                \hline
        120& -0.7015028275085&-0.8219288297026&-0.8806084906635&-0.9135413831900&-0.9472288106736\\
\hline
       240&-0.7023491359327&-0.8235146373945&-0.8826593378956&-0.9159126071869&-0.9500136623979\\
\hline
  600&-0.7028648366299&-0.8244749117138&-0.8838991109256&-0.9173448863563&-0.9516943959573\\
\hline

  $\infty$&-0.7032120767462&-0.8251189342374&-0.8847296926763&-0.9183039468245&-0.9528192495986\\
  \hline
        \end{tabular}
        \caption{Exact ground state energy per site of the $SU(N)$ Heisenberg chain with the fundamental irrep at each site, with open boundary conditions, as a function of the length L, obtained by  solving the Bethe ansatz equations of Eq. (\ref{BAE}). For $L=\infty$, we took the formula from [\cite{sutherland}].
          }
        \label{table_EGS_exact}
\end{table*}

 \section{DMRG results for the $SU(N)$ Heisenberg chain with open boundary conditions}
 \label{study}
 We have numerically investigated the Heisenberg $SU(N)$ chain with OBC with our DMRG algorithm for $N=3,4,5,6$ and $8$.
We have performed the infinite size DMRG until the desired length of the chain, $L$, was reached, and then we have performed some sweeps from left to right and from right to left through the finite-size DMRG. The implementation of the $SU(N)$ symmetry in the finite-size DMRG involves exactly the same steps as in the infinite part,
except that the left and the right blocks have now different sizes. In particular, the calculation of the matrix elements of the interaction term between the two blocks
will depend on the $S_L \subset S_{x}\otimes S_{L-x}$ {\it subduction coefficients} \cite{chen}, where $x$ is the length of the left block. These coefficients are computed using the very same method as the one developped in the previous part for the $S_L \subset S_{\frac{L}{2}}\otimes S_{\frac{L}{2}}$ {\it subduction coefficients}.
We have computed the ground state energies $\mathcal{E}_L$ at the end of the infinite size DMRG, and calculated the difference $\Delta_E^{\infty}$ with respect to the exact energies (shown in Table \ref{table_EGS_exact}). We have also calculated the ground state energy during the finite-size part of the DMRG. We have noticed that after two sweeps, $\mathcal{E}_L$ starts saturating. The corresponding difference with the exact energy, named $\Delta_E^{\text{Sw}}$ is typically one to two orders of magnitude smaller than  $\Delta_E^{\infty}$.
We list the results in Table (\ref{results_energies}) for $L=120$ and for $L=240$.
Note that similar comparisons have been performed for $SU(N)$ Hubbard chains, for which the Bethe ansatz is only approximate as soon as $N>2$\cite{manmana2011}.

 \label{results}
\begin{table}[h]
	\centering
	\begin{tabular}{|c|c|c|c|c|c|c|}
		\hline
		N&M&$\Delta_E^{\infty}(120)$&$\Delta_E^{\text{Sw}}(120)$&$\Delta_E^{\infty}(240)$&$\Delta_E^{\text{Sw}}(240)$&$\mathcal{W}_d^{120}$\\
		\hline
		3&85&$6\times10^{-11}$&$6\times10^{-13}$&$5\times10^{-11}$&$8\times10^{-13}$&$9\times10^{-12}$\\
		\hline
	4&160&$4\times10^{-9}$&$3\times10^{-10}$&$3\times10^{-9}$&$5\times10^{-10}$&$6\times10^{-9}$\\
\hline
	5&150&$1\times10^{-7}$&$4\times10^{-9}$&$3\times10^{-8}$&$8\times10^{-9}$&$1\times10^{-7}$\\
\hline
6&150&$1\times10^{-7}$&$1\times10^{-8}$&$9\times10^{-8}$&$3\times10^{-9}$&$2\times10^{-5}$\\
\hline
8&200&$9\times10^{-7}$&$7\times10^{-7}$&$4\times10^{-6}$&$4\times10^{-6}$&$2\times10^{-4}$\\
\hline
	\end{tabular}
	\caption{DMRG results for a chain of $L=120$ sites (third and fourth column) and of $L=240$ sites (fifth and sixth column), with open boundary conditions. $M$ is the number of irreps kept (cf section \ref{method}  for details),
	$\Delta_E^{\infty}$ is the difference between the ground state energy $\mathcal{E}_L$ obtained at the end of the infinite size part of the DMRG, and the exact Bethe ansatz energies $E(L)$ (shown in Table \ref{table_EGS_exact}). $\Delta_E^{Sw}$ corresponds to the same quantity after two sweeps, where $\mathcal{E}_L$ starts saturating. We have kept $m=1000$ states for  $L=120$ and $m=2000$ states for $L=240$.
	Seventh column: discarded weight $\mathcal{W}_d^{120}$ at the end of the infinite-size DMRG for $L=120$ and $m=1000$. }
	\label{results_energies}
\end{table}

\begin{figure} 
\centerline{\includegraphics[width=1\linewidth]{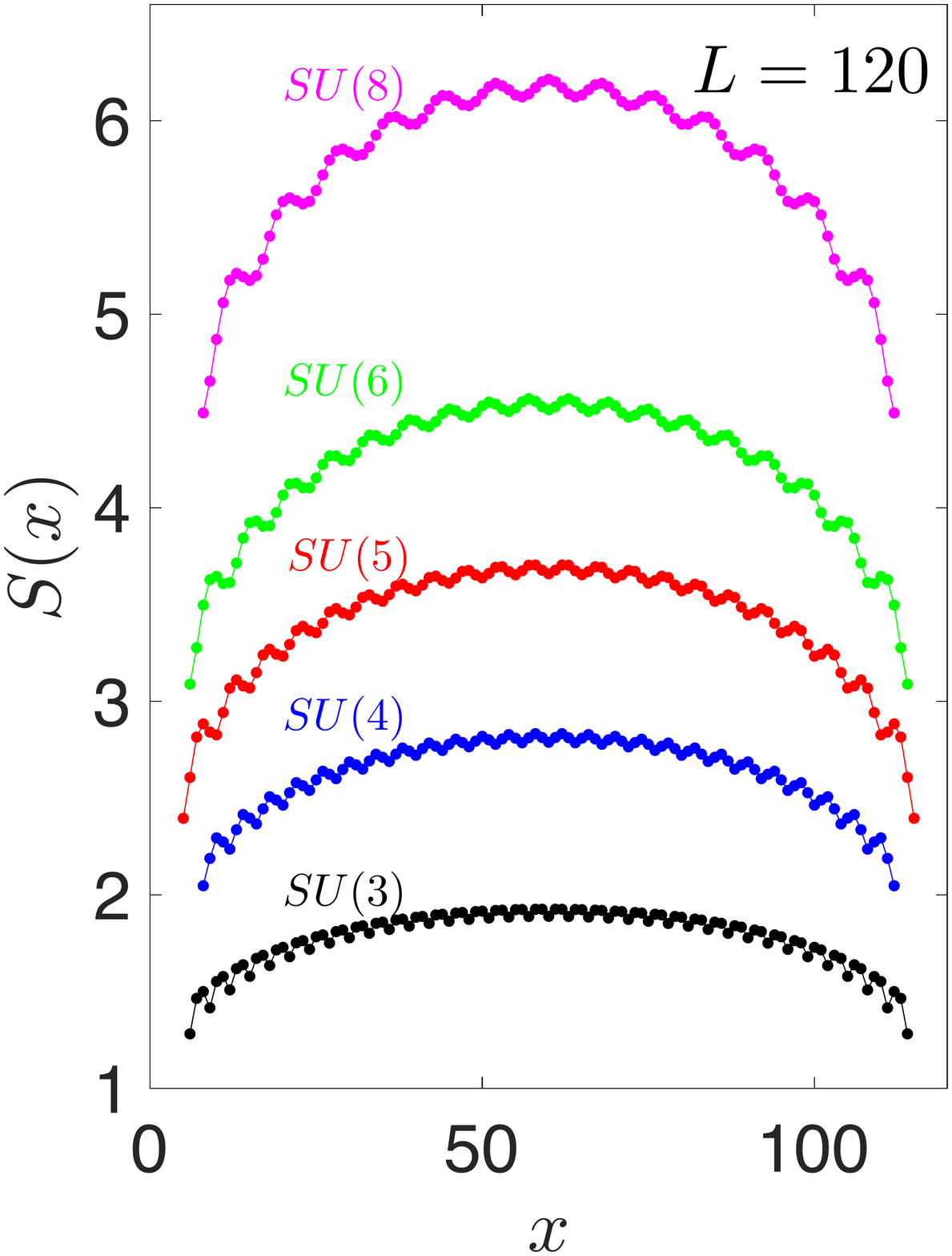}}
%\centerline{\includegraphics[width=1\linewidth]{entanglement_120sites}}
\caption{\label{entanglement_120}Entanglement entropy $S(x)$ as a function of the position $x$ along an open chain of $L=120$ sites.
The curves exhibit some N-periodic Friedel oscillations.}
\end{figure}

\begin{figure} 
\centerline{\includegraphics[width=1.05\linewidth]{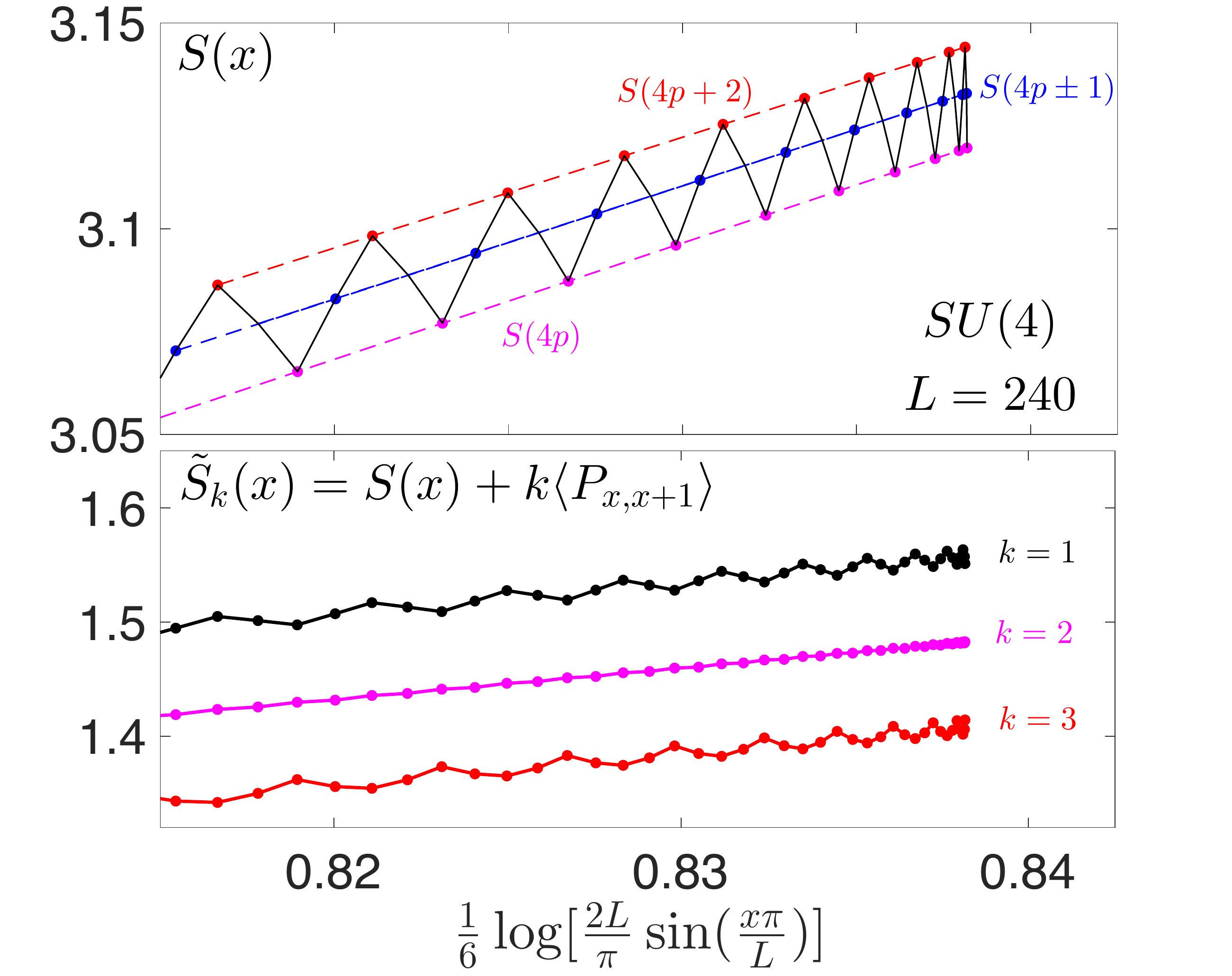}}
\caption{\label{stilde_k_SU4} Top: Entanglement entropy $S(x)$ as a function of the logarithm of the conformal distance $\frac{1}{6} \log\Big{[}\frac{2L}{\pi}\sin\Big{(}\frac{\pi x}{L}\Big{)}\Big{]}$, for an open chain of $L=240$ and $SU(4)$. The position $x$ can be written as $x=4\times p+q$ (with $q=0,1,2,3$), giving rise to three different straight lines: in magenta for $q=0$, in blue for $q=1,3$  or $q=\pm1$, and in red for $q=2$. The corresponding slopes are $c_{q}$: here, we have $c_{0}\approx 2.83$, $c_{1,3}=c_{\pm1}\approx 2.76$ and $c_2\approx 2.70$. Bottom: to remove Friedel oscillations, one can alternatively consider $\tilde{S}_k(x)=S(x)+k \langle P_{x,x+1}\rangle$ (see text for details). Here, the parameter $k=2$ allows one to  subtract the oscillations and lead to a straight line of slope $\tilde{c}\approx 2.75$. It is a general observation that $c_{N/2} \leq \tilde{c} \leq c_{0}$.}
\end{figure}

During the sweeps, we have also computed the entanglement entropy $S(x)$ (cf Eq. (\ref{entanglement_rho})) as a function of $x$, the position of the sweep (also the size of the left block). For critical spin chains with OBC, the entanglement entropy is given by the Calabrese-Cardy formula \cite{calabrese_cardy}:
\begin{align}
\label{calabrese_cardy}
S(x)=\frac{c}{6} \log\Big{[}\frac{2L}{\pi}\sin\Big{(}\frac{\pi x}{L}\Big{)}\Big{]}+K,
\end{align}
where $K$ is a non-universal constant, 
and c is the central charge of the associated CFT. In the case of the Heisenberg $SU(N)$ spin chains with the fundamental irrep at each site, the CFT is a Wess- Zumino-Witten model with topological coupling $k=1$ ($SU(N)_1$ WZW), with central charge $c_{\text{th}}=N-1$.
We show in Fig. \ref{entanglement_120}  the profile of the entanglement entropy $S(x)$ for $L=120$,
as a function of the position $x$.
Because of the PBC, $S(x)$ has Friedel oscillation with $N$-fold periodicity.
%We also show on these figures $S(x)$ as a function of $(1/6)\log\Big{[}(2L/\pi)sin\Big{(}\pi x/L\Big{)}\Big{]}$, for $x$ multiple of $N$ in order to avoid the oscillations.
%According to Eq. (\ref{calabrese_cardy}), these latter curves should correspond to some straight lines with a slope equal to the central charge.
%We have fitted such slopes, called $c_{\text{f}}$, and we report the corresponding values in Table \ref{results_120} and \ref{results_240}.
%The accuracy ($\pm 0.01$ for instance) in the column for  $c_{\text{f}}$ is a indication of how well Eq. (\ref{calabrese_cardy}) is satisfied for $S(x)$.
In order to remove the oscillations and to fit the central charge, we have adopted two strategies, illustrated in Fig. \ref{stilde_k_SU4}, where we analyse the case of $SU(4)$,
$L=240$.
Since the oscillations are $N-$periodic, one can plot $S(x)$ as a function of the logarithm of the conformal distance $\frac{1}{6} \log\Big{[}\frac{2L}{\pi}\sin\Big{(}\frac{\pi x}{L}\Big{)}\Big{]}$ separately for different sets of abscissa $x$
%which have the same remainder in the Euclidean division by $N$, i.e. for sets indexed by $q$ 
of the form $x=N\times p+q$. Each set corresponds to a fixed $q=0,1,..N-1$ and to all values of $p$ consistent with the overall length.
It then gives rise to $N$ different straight lines (one for each $q$), as shown in top of Fig. \ref{stilde_k_SU4}, with different slopes $c_q$. It turns out that one always has $c_{N/2} \leq ...\leq c_{0}$. In particular, fitting $S(x)$ for $x$ multiple of $N$ always gives rise to the largest central charge, while fitting $S(x)$ for $x\equiv N/2$ mod $N$ always gives the smallest one.
Note that when $N$ is odd, $N/2$ should be understood as $\text{Floor}(N/2)$.
Alternatively, one can also follow the strategy developed in \cite{Capponi_oscillations_2013}: Since the Friedel oscillations originate from the bond modulations, it is convenient to introduce $\tilde{S}_k(x)=S(x)+k \langle P_{x,x+1}\rangle$,
where $\langle P_{x,x+1}\rangle$ is the expectation value on the ground state of the bond operator $ P_{x,x+1}$, and where $k$ is a parameter that is adjusted to best remove the oscillations.
As shown in the bottom panel of Fig. \ref{stilde_k_SU4}, it appears that for $SU(4)$, one should take $k=2$.
 The best value of $k$ is $N-$dependent but does not change with $L$. The best parameters $k$ were thus found to be $k=1,2,3.5$ and $6$ for respectively $N=3,4,5$ and $6$. 
For theses values of $k$, we then fit $\tilde{S}_k(x)$ as a function of the logarithmic of the conformal distance to obtain $\tilde{c}$. We have systematically observed that $c_{N/2} \leq \tilde{c} \leq c_{0}$.

\begin{figure} 
\centerline{\includegraphics[width=1.15\linewidth]{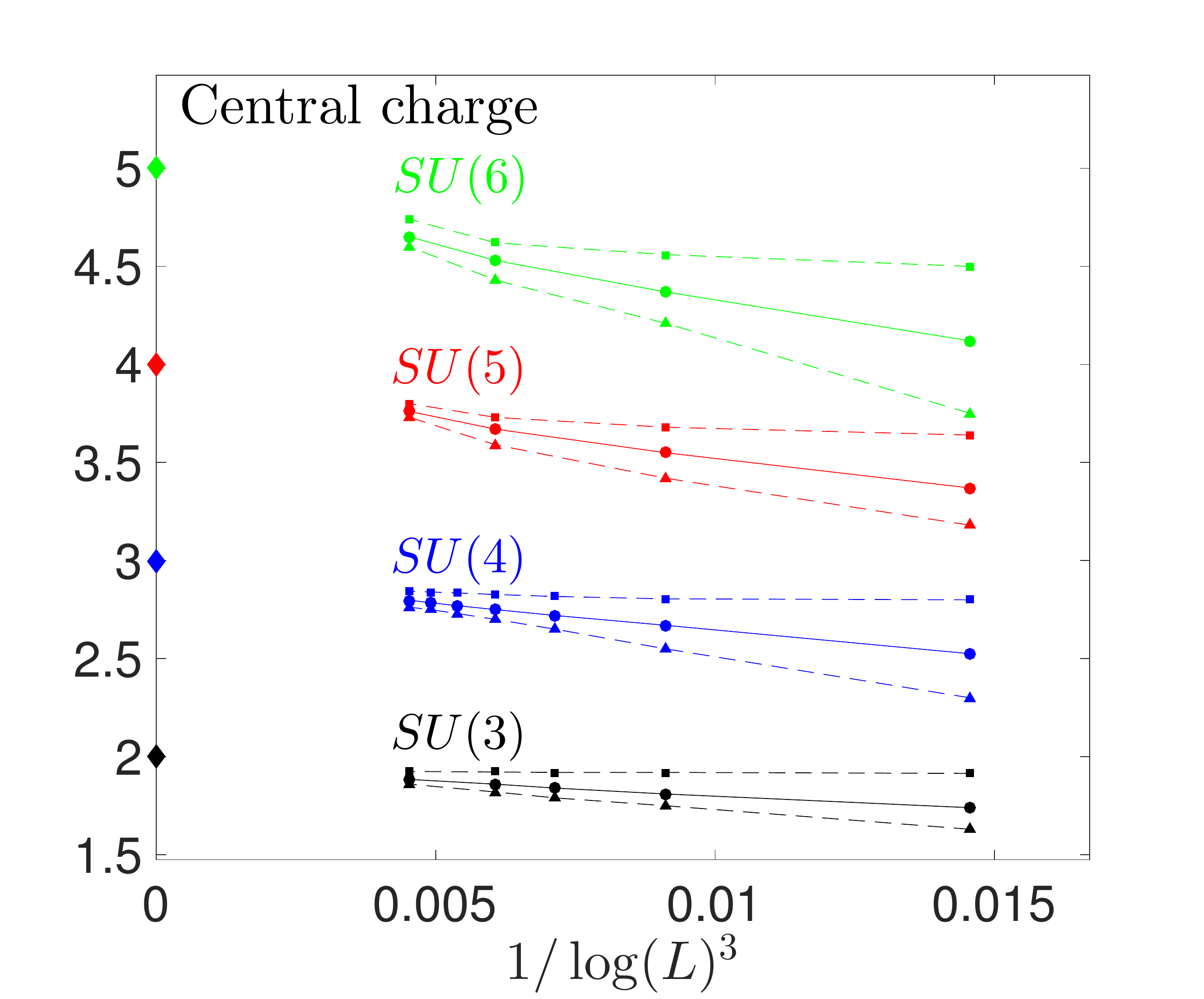}}
\caption{\label{central_charges} Fitted central charges for $SU(3)$ up to $SU(6)$ from $L=60$ to $L=420$, as a function of $1/\log(L)^3$:
$\tilde{c}$ (solid circles), $c_0$ (squares) and $c_{N/2}$ (triangles). These estimates of the central charges, all based on the DMRG calculation of the entanglement entropy, are defined in the main text.
Diamonds correspond to the theoretical $SU(N)_1$ central charges: $c_{\text{th}}=N-1$.
The finite-size central charges are consistent with the theoretical values in the thermodynamical limit.}
\end{figure}

In Fig. \ref{central_charges}, following Ref. [\onlinecite{ziman1987}], we plot  the quantities $c_{N/2}$, $\tilde{c}$ and $c_{0}$ as a function of $1/\log(L)^3$ for different values of $N$.
%From a physical point of view, these results demonstrate some relative good agreement with the expected $SU(N)_1$ CFT.
%Our fitted central charges converge towards the theoretical values $c_{\text{th}}=N-1$ expected for $SU(N)_1$.
The finite-size effects are large, but the results are consistent with the theoretical values $c_{\text{th}}=N-1$ expected for $SU(N)_1$ in the thermodynamic limit. For instance, for $SU(5)$, $\tilde{c}\approx 3.37$ for $L=60$ while for $L=420$, we have $\tilde{c}\approx 3.76$, much closer to the thermodynamical limit $c_{\text{th}}=4$. Note that, as for $SU(2)$, the finite-size effects are much stronger than with periodic boundary conditions, as revealed by a comparison with the DMRG results of Ref. [\onlinecite{fuhringer2008}] for SU(3) and SU(4). Work is in progress to extend our algorithm to periodic boundary conditions. 

 \section{Conclusions and Perspectives}
To conclude, we have presented an efficient DMRG algorithm to study the $SU(N)$ Heisenberg model that takes into account the full $SU(N)$ symmetry. The efficiency of this algorithm relies on a systematic use of the orthogonal representation of the symmetric group to shortcut the calculation of many $SU(N)$ group theory coefficients needed in the computation.
In particular, we have shown how to calculate the matrix elements of the inter-block interaction term from the permutational subduction coefficients. 
The calculation of these coefficients is essentially based on Chen's method, with an improvement that makes the calculation for larger $N$ much lighter.
%To calculate these  coefficients,  we have first used Chen's method \cite{chen} which is based on SYT and the Young's rules, and then proposed an improvement of this latter that renders calculation for larger $N$ much easier.
We have applied our method to calculate the ground state energies and the entanglement entropies of the Heisenberg $SU(N)$ chains with one particle per site and with OBC for several hundreds of sites.
Our results agree well with both the exact $SU(N)$ ground state energies obtained from Bethe ansatz (in section \ref{exact}), and with the central charges associated to the expected $SU(N)_1$ CFT. They also point to significant finite-size effects when calculating the central charge. The magnitude of these finite-size effects increases with $N$ for a given length of the chain, calling for a better and more systematic understanding of the finite size effects in the future.

From a methodological point of view, we plan to make several developments that are related to relevant open physical questions.

First of all, we would like to implement the calculation of the subduction coefficients when the outer multiplicity is strictly larger than 1.
 This will enable us to build the matrix representing the Hamiltonian of the entire chain on a target irrep that does not necessarily contain the ground state, and thus to access
 some excited states of the full Hilbert space which are states of minimal energy in their sector.
 
 Secondly, we would like to generalize the procedure to situations where the irrep on each site is not the fundamental one anymore.
 For instance, we would like to study the case where the irrep on each site is the fully symmetric one in order to check the generalization of the Haldane conjecture to $SU(N)$.
 Numerically, this problem has already been studied for $N=3,4$ \cite{rachel2009} but many questions remain open (including for $SU(3)$ and $SU(4)$)\cite{PRBNataf2016}. 
 Following the recent analytical filed theory approach \cite{Lajko_2017}, the $SU(3)$ Heisenberg chain with the fully symmetric irrep $\ydiagram{3}$ on each site should be gapped.
 A natural question to be addressed by DMRG is for instance the value of this gap in the thermodynamical limit.
 Moreover, since our DMRG algorithm keeps track of the $SU(N)$ irrep of each state, we could take advantage of it to characterize the non-trivial edge states structure appearing in the case of OBC.
 One could also study the case of antisymmetric irreps on each site. For instance, the $SU(6)$ Heisenberg chain with two particles per site in the irrep $[1,1]$ is believed to be gapped \cite{Affleck1988,dufourPRB2015},
but the calculation of this gap still needs to be done. 
%Moreover, the nature of the phase for the SU(8) chain with irrep $[1,1]$ is still unknown.
In addition, with mixed irreps at each site, one could also investigate more sophisticated models such as the $SU(N)$ version of the AKLT models \cite{AKLT1,rachel2007,Morimoto2014,Wan2017} .
 They involve irreps with multiple rows and multiple columns at each site and can lead to N distinct topological phases\cite{nonne2013,Capponi_annals_2016}, classified using group cohomology\cite{Quella2013_phases,quella2012}.
 
Finally, Young's rules could also be used to implement $SU(N)$ symmetries in tensor networks to study two-dimensional $SU(N)$ Heisenberg models, which could potentially host exotic physical phases \cite{hermele2009,hermele_topological_2011,Lecheminant2016,Lecheminant2016b,Dufour2016,natafPRL2016,natafrapidcomm2016,Kim_2017}.
  %These models can host exotic physical phases: for one particle per site, they can give rise to abelian chiral spin liquids described by the $SU(N)_1$ WZW CFT \cite{Dufour2016,natafPRL2016,natafrapidcomm2016}.
  %With several particle per site and irrep of  mixed symmetry, this class of models could potentially host  non-abelian chiral spin liquids described by the $SU(N)_k$ WZNW CFT (with $k > 1$) .
%  Finally, one could try to generalize the procedure to other models such as the $SU(N)$ Fermi-Hubbard or to models invariant under other groups of symmetry, such as the $Sp(N)$ \cite{SpN_kataoka,coleman_spn, SPN_ramires} or $SO(N)$-invariant models.
Work is currently in progress along these lines.
 
 \section{Acknowledgements} 
We acknowledge useful discussions with Jean-S\'ebastien Caux, Natalia Chepiga, Luc Frappat and Wen-Li Yang.
This work has been supported by the Swiss National Science Foundation. The calculations have been performed using the facilities of the Scientific IT and Application Support Center of EPFL.
  
 \section{Appendix}
\label{Appendix}
\subsection{ Irreps of $\mathrm{SU}(N)$ }
\label{Appendix_irrep}

In general, for a system of $n$ particles, each irrep of $\mathrm{SU}(N)$ can be associated with a Young diagram composed of $n$ boxes arranged in at most $N$ rows. This represents a particular set of $\mathrm{SU}(N)$-symmetric $n$-particle wave functions. The shape $\alpha$ of the Young diagram is specified by a partition $\alpha = [\alpha_{1}, \alpha_{2}, \dots,\alpha_{k}]$ (with $1 \leq k \leq N$ and $\sum_{j=1}^{k} \alpha_{j} = n$), where the row lengths $\alpha_{j}$ satisfy $\alpha_{1} \geq \alpha_{2} \geq \dots \geq \alpha_{k} \geq 1$. The diagram can be filled with numbers from 1 to $n$, and the resulting tableau is said to be \textit{standard} if the entries are increasing from left to right in every row and from top to bottom in every column. Standard Young tableaux (SYTs) play a central role in the representation theory of the symmetric group.

Using $[1]=\Box$ to denote the fundamental irrep, the set of all $n$-particle wave functions live in the full Hilbert space $\Box^{\otimes n}$. The multiplicity $f^{\alpha}$ of irrep $\alpha$ in this space is equal to number of SYTs with shape $\alpha$. This number can be calculated from the hook length formula,
\begin{equation}
f^{\alpha} = \frac{n!}{\prod_{i=1}^{n}l_{i}},
\end{equation}
where the hook length $l_{i}$ of the $i$-th box is defined as the number of boxes to the right of it in the same row, plus the number of boxes below it in the same column, plus one (for the box itself). The dimension $d_{N}^{\alpha}$ of the irrep can also be calculated from the shape as
\begin{equation} \label{E: dN}
 d_{N}^{\alpha} = \prod_{i=1}^{n}\frac{N+\gamma_{i}}{l_{i}},
\end{equation}
where $\gamma_{i}$ is the algebraic distance from the $i$-th box to the main diagonal, counted positively (resp. negatively) for each box above (resp. below) the diagonal. The full Hilbert space can be decomposed as 
\begin{equation} \label{E: sectors}
\Box^{\otimes n} = \oplus_{\alpha} V^{\alpha},
\end{equation} where $V^{\alpha}$ is the sector corresponding to irrep $\alpha$ (and if $d_{N}^{\alpha} > 1$, $V^{\alpha}$ can itself be decomposed into $d_{N}^{\alpha}$ equivalent subsectors, $V^{\alpha}= \oplus_{i=1}^{d^{\alpha}_N} V^{\alpha}_i $).
The equation for the dimension of the full Hilbert space thus reads
\begin{align}
\label{dimension_full}
N^{n}=\sum_{\alpha} f^{\alpha} d^{\alpha}_N,
\end{align}
where the sum runs over all Young diagrams with $n$ boxes and no more than $N$ rows.

As an example, for $n=2$, $\Box^2$ can be decomposed as a sum of the subspace spanned by the symmetric two-particle wave functions and of that spanned by the antisymmetric two-particle wave functions: 
\[ \ytableausetup{smalltableaux}\ydiagram{1}^{\otimes 2} = \ydiagram{1} \otimes \ydiagram{1}= \ydiagram{2}  \oplus \ydiagram{1,1}. \]
There is only one SYT for each of the diagrams [2] and [1,1], so $f^{[2]}=f^{[1,1]}=1$, while Eq. \eqref{E: dN} gives $d_{N}^{[2]}=\frac{N(N+1)}{2}$ and $d_{N}^{[1,1]}=\frac{N(N-1)}{2}$. It is easy to check that Eq. (\ref{dimension_full}) is satisfied.

\subsection{Young's orthogonal representation of the symmetric group} \label{young_rep}
This subsection summarizes some useful results concerning the orthogonal representation of the symmetric group.

For a given tableau shape $\alpha$, a convenient representation of the symmetric group $\mathcal{S}_{n}$ can be formulated using Young's \textit{orthogonal units} $\{o_{rs}^{\alpha}\}_{r,s=1\dots f^{\alpha}}$. These are specific linear combinations of permutations, whose explicit forms are not needed here (but could be explicitely calculated \cite{rutherford,Wan2017}). They satisfy orthonormality:
\begin{equation}
o^{\alpha}_{rs}o^{\beta}_{uv} = \delta^{\alpha\beta}\delta_{su}o_{rv}^{\alpha} \quad \forall r,s = 1\dots f^{\alpha}, \forall u,v = 1\dots f^{\beta}
\label{orthonormal_relation} 
\end{equation}
as well as completeness:
\[ %\begin{equation}
\sum_{\alpha} \sum_{r=1}^{f^{\alpha}} o^{\alpha}_{rr} = Id
\label{completeness_formula} 
\] %\end{equation}
and form a basis in which any linear superposition $\eta$ of permutations belonging to $\mathcal{S}_{n}$ can be uniquely decomposed as
\begin{equation}
\label{E: unique_decomposition}
 \eta = \sum_{\alpha,r,s} \mu_{rs}^{\alpha}(\eta)o_{rs}^{\alpha}, 
\end{equation}
where $\mu_{rs}^{\alpha}(\eta)$ are real coefficients.

An important result we will make frequent use of is that successive transpositions $P_{k,k+1}$, i.e. permutations between the consecutive numbers $k$ and $k+1$ ($1 \leq k \leq n-1$), takes an extremely simple form in the basis of orthogonal units. If we write $P_{k,k+1} = \sum_{\alpha,t,q} \mu_{tq}^{\alpha}(P_{k,k+1})o_{tq}^{\alpha}$, then, for a given shape $\beta$, the matrices $\bar{\bar{\mu}}^{\beta}(P_{k,k+1})$ defined by\[\left[\bar{\bar{\mu}}^{\beta}(P_{k,k+1})\right]_{tq} = \mu_{tq}^{\beta}(P_{k,k+1})\] are symmetric and orthogonal, and very sparse, with at most two nonzero entries in each row and in each column. These entries can be explicitly calculated as follows. We first assign some fixed order (named {\it last letter order}) to the $f^{\alpha}$ SYTs :
two standard tableaux $S_r$ and $S_s$ are such that $S_r < S_s$ if the number $n$ (which is the last one) appears in $S_r$ in a row below the one in which it appears in $S_s$. If those rows are the same, one looks at the rows with $n-1$ , etc.
Thus,  we can label the SYTs  $S_{1},\dots,S_{f^{\beta}}$.
 Then, if $k$ and $k+1$ are in the same row (resp. column) in the tableau $S_{t}$, we will have $\mu_{tt}^{\beta}(P_{k,k+1}) = +1$ (resp. $-1$), and all other matrix elements involving $t$ will vanish. If $k$ and $k+1$ are not in the same column nor the same row in $S_{t}$, and if $S_{q}$ is the tableau obtained from $S_{t}$ by interchanging $k$ and $k+1$, then the only non-vanishing matrix elements involving $t$ or $u$ are given by
\begin{equation} \label{E: Y_rules}
\begin{pmatrix} \mu_{tt}^{\beta}(P_{k,k+1}) \enspace & \mu_{tq}^{\beta}(P_{k,k+1}) \\[0.1cm] \mu_{qt}^{\beta}(P_{k,k+1}) \enspace & \mu_{qq}^{\beta}(P_{k,k+1}) \end{pmatrix} = \begin{pmatrix} -\rho \enspace & \sqrt{1-\rho^{2}} \\ \sqrt{1-\rho^{2}} \enspace & \rho \end{pmatrix}
\end{equation}
Here, $\rho$ is the inverse of the \textit{axial distance} from $k$ to $k+1$ on $S_{t}$, which is computed by counting $+1$ (resp. $-1$) for each step made downward or to the left (resp. upward or to the right) to reach $k+1$ from $k$. 

These simple yet incredibly useful formulas (that we call {\it Young rules}) are clarified in Figure \ref{F: Y_rules}. Moreover, since every permutation can be factorized into successive transpositions, we can use these rules to write the exact matrix representation of any permutation or linear superposition thereof via a few elementary calculations.

\begin{figure} \label{F: Y_rules}
\centerline{\includegraphics[width=0.5\linewidth]{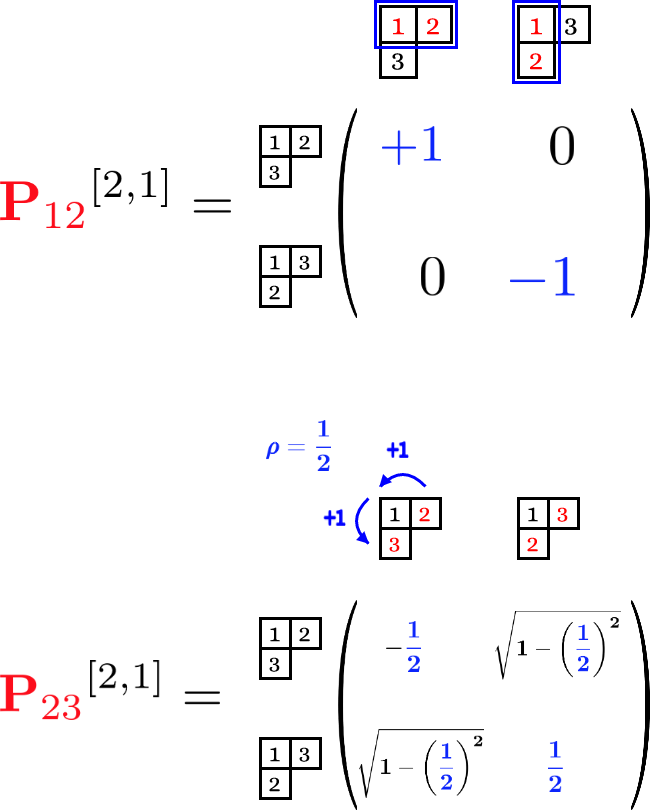}}
\caption{Writing the matrix representations of permutation operators $P_{12}$ and $P_{23}$ in the basis of SYTs of shape $\alpha = [2,1]$. We have labelled $\ytableaushort{12,3} = S_{1}$ and $\ytableaushort{13,2} = S_{2}$. For $P_{12}$, the numbers $1$ and $2$ are in the same row on $S_{1}$ and in the same column on $S_{2}$. For $P_{23}$, the axial distance between $1$ and $2$ on $S_{1}$ is 2, so $\rho = \frac{1}{2}$. $S_{2}$ is the tableau obtained from $S_{1}$ by interchanging $1$ and $2$, and we apply Eq. \eqref{E: Y_rules} accordingly. \label{F: Y_rules}}
\end{figure}

\subsection{Calculation of the IRF weights in the SU(2) case using the SYTs and the Young's rules}
\label{irf}
We calculate here the IRF weights for the SU(2) Heisenberg model, which appear in Table 2 p 520 of \cite{Sierra_1996}. Contrary to what has been done in  \cite{Sierra_1996}, we do not perform this calculation using the 6j-symbols (cf Eq. (62) of \cite{Sierra_1996}), but rather using the SYTs and the Young's rules, which are much simpler.
First, we can establish a connection between the notations appearing in Fig \ref{chain} and in the section \ref{construction_left} of the present manuscript and the ones which appear in Table 2 p 520  of \cite{Sierra_1996}:
\begin{align}
\chi  &\leftrightarrow a \\ \nonumber
\beta &\leftrightarrow c \\ \nonumber
\alpha'&\leftrightarrow d\\ \nonumber 
\alpha &\leftrightarrow b \nonumber
\end{align}

The coefficient:
\begin{align}
R\begin{pmatrix} & d & \\ a & & c \\ & b &
\end{pmatrix}
\end{align}
is equal to the matrix element of $P_{N_s,N_s+1}$ between a state $\vert\eta\rangle$ and a state $\vert\xi\rangle$, which both live in the irrep $c$,
and which were both built from a common "grandfather"   (that one can call $\vert \Psi\rangle$) by adding twice the fundamental irrep (spin $1/2$), corresponding to site $N_s$ and site $N_s+1$.
The common "grandfather" $\vert \Psi \rangle $ is in the irrep $a$. For the state $\vert \eta \rangle$, after adding the first fundamental irrep, the resulting state ("$\vert \eta^{-} \rangle$"), who is the direct ascendant (or "father") of $\vert \eta\rangle$, lives in the irrep $b$, while the father of $\vert \xi \rangle$  ("$\vert \xi^{-} \rangle$") lives in the irrep  $d$ .
One can also use chains of irreps to characterize these two states:
\begin{align}
\vert \eta \rangle &:\vert \Psi\rangle  \in a\rightarrow \vert \eta^{-} \rangle \in b \rightarrow \vert \eta \rangle \in c \,\,  \text{or }    \vert \eta \rangle : a\rightarrow b \rightarrow c \nonumber \\
\vert \xi \rangle &:\vert \Psi\rangle  \in a\rightarrow \vert \xi^{-} \rangle \in d \rightarrow \vert \xi \rangle \in c  \,\, \text{or }  \ \vert \xi \rangle :  a\rightarrow d \rightarrow   c  \nonumber
\end{align}
Now, for SU(2), one can label the irreps either by the value of the total spin $j$ or by using Young diagrams  $[\alpha_1,\alpha_2]$, where $\alpha_1-\alpha_2=2j$.\\
Thus the coefficient 
\begin{align}
R\begin{pmatrix} & j+1/2 & \\ j & & j \\ & j+1/2 &
\end{pmatrix} 
\end{align}
is equal to:
\begin{widetext}
\begin{align}
 \raisebox{-2ex}{\scalebox{4}{$\langle$}} \scalebox{3}{\begin{ytableau}  \,&  \none[\scalebox{0.5}{$\cdots$}]& \,  & \, &  \none[\scalebox{0.5}{$\cdots$}] \, & \,\scalebox{0.3} {$N_s$}\ \\  \,&  \none[\scalebox{0.5}{$\cdots$}] & \scalebox{0.25} {$N_s+1$}\end{ytableau}}   \raisebox{-2ex}{\scalebox{4}{$\vert$}} \scalebox{1.5}{$P_{N_s,N_s+1}$}  \raisebox{-2ex}{\scalebox{4}{$\vert$}}  \scalebox{3}{\begin{ytableau}  \,&  \none[\scalebox{0.5}{$\cdots$}]& \,  &\, &  \none[\scalebox{0.5}{$\cdots$}]& \, \,\scalebox{0.3} {$N_s$}\ \\  \,&  \none[\scalebox{0.5}{$\cdots$}] & \scalebox{0.25} {$N_s+1$}\end{ytableau}} \raisebox{-2ex}{\scalebox{4}{$\rangle$}} =\frac{-1}{2j+1},
\end{align}
\end{widetext}
where we have just used the rule described in the previous section about the diagonal term of the matrix representing $P_{N_s,N_s+1}$, with $2j+1$ the axial distance between
$N_s$ and $N_s+1$ since the young diagram is such that $\alpha_1-\alpha_2=2j$.
It also directly follows that:
\begin{widetext}
\begin{align}
R\begin{pmatrix} & j+1/2 & \\ j & & j \\ & j-1/2 &
\end{pmatrix} =& \raisebox{-2ex}{\scalebox{4}{$\langle$}} \scalebox{3}{\begin{ytableau}  \,&  \none[\scalebox{0.5}{$\cdots$}]& \,  & \, & \none[\scalebox{0.5}{$\cdots$}]& \, \,\scalebox{0.3} {$N_s$}\ \\  \,&  \none[\scalebox{0.5}{$\cdots$}] & \scalebox{0.25} {$N_s+1$}\end{ytableau}}   \raisebox{-2ex}{\scalebox{4}{$\vert$}} \scalebox{1.5}{$P_{N_s,N_s+1}$}  \raisebox{-2ex}{\scalebox{4}{$\vert$}}  \scalebox{3}{\begin{ytableau}  \,&  \none[\scalebox{0.5}{$\cdots$}]& \,  &  \, &\none[\scalebox{0.5}{$\cdots$}]& \,  \scalebox{0.25} {$N_s+1$}\ \\  \,&  \none[\scalebox{0.5}{$\cdots$}] & \scalebox{0.3} {$N_s$}\end{ytableau}} \raisebox{-2ex}{\scalebox{4}{$\rangle$}} \nonumber \\ 
&=\sqrt{1-(\frac{1}{2j+1})^2}=\frac{\sqrt{2j(2j+2)}}{2j+1},
\end{align}
\end{widetext}
using the rule for the off-diagonal term of the matrix representing $P_{N_s,N_s+1}$, with $2j+1$ the axial distance between
$N_s$ and $N_s+1$.
One also finds:
\begin{widetext}
\begin{align}
&R\begin{pmatrix} & j & \\ j-1/2& & j +1/2\\ & j &
\end{pmatrix} =\nonumber \\& \raisebox{-2ex}{\scalebox{4}{$\langle$}} \scalebox{3}{\begin{ytableau}  \,&  \none[\scalebox{0.5}{$\cdots$}]& \,  & \, & \none[\scalebox{0.5}{$\cdots$}]& \, \scalebox{0.3} {$N_s$}& \scalebox{0.25} {$N_s+1$} \\  \,&  \none[\scalebox{0.5}{$\cdots$}] & \, \end{ytableau}}   \raisebox{-2ex}{\scalebox{4}{$\vert$}} \scalebox{1.5}{$P_{N_s,N_s+1}$}  \raisebox{-2ex}{\scalebox{4}{$\vert$}}  \scalebox{3}{\begin{ytableau}  \,&  \none[\scalebox{0.5}{$\cdots$}]& \,  & \, & \none[\scalebox{0.5}{$\cdots$}]& \, \scalebox{0.3} {$N_s$} & \scalebox{0.25} {$N_s+1$} \\  \,&  \none[\scalebox{0.5}{$\cdots$}] & \, \end{ytableau}} \raisebox{-2ex}{\scalebox{4}{$\rangle$}} \nonumber \\ 
&=1.
\end{align}
\end{widetext}
The other coefficients can be calculated in exactly the same way.
We see that the Young's rules allow one to completely bypass the calculation of the $6j-symbols$ to obtain the IRF-weights.
For spins $1$, the IRF weights obtained by Tatsuaki \cite{tatsuaki} can also be easily calculated using SYTs and the rules we developed for the case of purely symmetric irrep\cite{PRBNataf2016}.
Finally, we want to point out that the SU(2) case is very specific: all the coefficients needed for the DMRG can be expressed in terms of IRF weights.
In particular, the interblock interaction can be easily written in term of the IRF weights in the Nishino and Sierra's approach. The underlying reason is that for $SU(2)$,
an irrep is equal to its conjugate. Such a fact manifests itself into some symmetries between the  $6j-symbols$.

\bibliography{DMRG.bib}
\end{document}